\documentclass[useAMS,usenatbib]{mn2e}
\usepackage{amsmath}
\usepackage{amsbsy}
\usepackage{times}
\usepackage{graphicx}
\usepackage{aas_symbols}
\usepackage{epstopdf}
\usepackage{cases}
\usepackage{color}
\usepackage{url}

\begin{document}
	
	\title[Constraints on GRB jet structure]{Observational constraints on the structure of gamma-ray burst jets}
	
	\author[Beniamini $\&$ Nakar]{Paz Beniamini$^{1}$\thanks{Email: paz.beniamini@gmail.com}, Ehud Nakar$^{2}$\\
		$^{1}$Department of Physics, The George Washington University, Washington, DC 20052, USA \\	
		$^{2}$The Raymond and Beverly Sackler School of Physics and Astronomy, Tel Aviv University, Tel Aviv 69978, Israel}
	
	\date{Accepted; Received; in original form ...}
	
	\pubyear{2019}
	
	\maketitle
	\begin{abstract}
		Motivated by GW170817 we examine constraints that observations put on the angular structure of long gamma-ray burst (GRB) jets. First, the relatively narrow observed distribution of $E_{X}/E_{\gamma}$ (the isotropic equivalent early X-ray afterglow to prompt $\gamma$-ray energy ratio) implies that at any angle that $\gamma$-rays are emitted the Lorentz factor must be high. Specifically, the Lorentz factor of $\gamma$-ray emitting material cannot drop rapidly with angle, and must be $\Gamma(\theta)\gtrsim 50$ even if there are angles for which the gamma-ray received energy is lower by three orders of magnitude compared to the jet core. Second, jets with an angular structure of the $\gamma$-ray emission that over-produce events with a $\gamma$-ray luminosity below the peak of the observed luminosity function are ruled-out. This eliminates models in which the $\gamma$-ray energy angular distribution isn't sufficiently steep and the Lorentz factor distribution isn't sufficiently shallow. Finally, models with a steep structure (e.g. Gaussian) which are detected away from the jet core generate afterglow light-curves that were never observed. We conclude that even if the jet kinetic energy distribution drops continuously with latitude, efficient $\gamma$-ray emission seems to be restricted to material with $\Gamma\gtrsim 50$ and is most likely confined to a narrow region around the core. While our study is confined to long GRBs, where the observed sample is larger and more complete, there are indications that similar conclusions may be applicable also to short GRBs. We discuss the possible implications to the $\gamma$-rays observed in GRB 170817A.
	\end{abstract}
	
	
	\section{Introduction}
	In August 2017, GW 170817 the first gravitational wave (GW) signal from a neutron star - neutron star merger was detected \citep{GW170817}. The accompanying radio and X-ray afterglow has been observed extensively over a period of almost a year \citep[e.g.,][]{Hallinan2017,Troja2017,Margutti2017,Ruan2018,Mooley2018}. The latest turnover in the radio and X-ray light-curves, as well as the measurement of superluminal motion from the radio data, suggests the existence of a successful relativistic jet and a structure with a significant angular energy profile in which we are observing the burst at an angle that is at least three times larger than the jet's opening angle \citep{Mooley2018B}. There still however, remains the question of whether the prompt gamma-rays are produced by radiatively efficient (but less energetic due to a steep angular structure) material at the `wings' of the jet or by the a shock breakout from the cocoon generated by the jet-ejecta interaction (in which case the energy content at the observed viewing angle is larger, but the efficiency smaller).

	In light of this, the allowed structure of GRB jets and the efficiency at which it produces $\gamma$-rays at large angles remains a topic of major importance, and it is useful to consider what types of jet structures are consistent with GRB observations \citep[see also][]{Beniamini2018B}. Previous studies have considered the implications of structure models on the true energetics and rates of GRBs \citep{Lipunov2001,Frail2001,Rossi2002,Zhang2002,Eichler2004,VanEerten2012,Pescalli2015}, on the shape of the afterglow light-curve \citep{KG2003,GK2003,Salmonson2003} or on detectability of orphan afterglows \citep{Lamb2017}. Here we propose a novel way to test the allowed structure of GRBs (in terms of both the energy and Lorentz factor angular distributions), by applying three independent techniques. We focus on long GRBs for which more detailed observations are available. First, we compare the predictions of these models regarding the $E_{X}/E_{\gamma}$ distribution (i.e. the isotropic equivalent early X-ray afterglow to prompt $\gamma$-ray energy ratio) to the observations. We show that a variety of structure models predict large variations in this quantity, in contrast with results from GRB observations. Secondly, we reconsider the effect of the structure on the observed luminosity function and show that a large family of models can be ruled out as they lead to an overproduction of bursts with $\gamma$-ray luminosities below the peak of the observed luminosity function. Both these considerations imply that while the energy angular profile may be steep, the Lorentz factor of GRBs must remain large at any region that produces $\gamma$-rays efficiently. However, even such models typically lead to very peculiar light-curves that can be ruled out by observations.
	The most likely implication is that efficient $\gamma$-ray emission must be confined to a narrow opening angle around the jet's core, where the isotropic equivalent energy is not much lower than that of the core. This will naturally resolve all the problems mentioned above.

	The paper is organized as follows. In \S \ref{sec:obs} we describe the observational results regarding the $E_{X}/E_{\gamma}$ distribution. We then describe in \S \ref{sec:model} the structure models and their resulting emission in the prompt $\gamma$-ray and in the X-ray afterglow and derive approximate limits on the Lorentz factor at different observation angles. In \S \ref{sec:results} we discuss our results, comparing the different models with the scatter observations. In \S \ref{sec:lumfunc} we turn to examine the constraints imposed by the observed luminosity function. We explore the possibility that $\gamma$-ray production is restricted to a narrow range beyond the core in \S \ref{narrowgamma}. We then conclude in \S \ref{conclusions}.
	\section{$L_X/E_{\gamma}$ (and $E_X/E_{\gamma}$) Observations}
	\label{sec:obs}
	In order to compare the results of different structure models with observations we select a sample of GRBs with measured (isotropically equivalent) energy emitted in X-rays $L_X$ and (isotropically equivalent) energy released during the prompt phase $E_{\gamma}$.
	We use the BAT6 sample, a sample of long Swift GRBs that were chosen to be almost complete in redshift (for details see \citealt{Salvaterra2012}). The required information on $L_X$ and $E_{\gamma}$ is available for 43 bursts in the BAT6 sample. The values of $L_X$ (integrated in the rest frame energy range 2-$10\,$keV and measured at different source frame times after the trigger) can be found in Table 1 of \cite{D'Avanzo2012} and the values of the prompt energies $E_{\gamma}$ are reported in \cite{Nava2012}. The best linear fits between $L_X$ (estimated at different times) and $E_{\gamma}$ are given by (see also \citealt{Beniamini2016}):
	\begin{eqnarray}
	\label{eq:correl}
	&L_{X,45}\!=\!120\,E_{\gamma,52}\quad\sigma_{log (L_{X}/E_{\gamma})}\!=\!0.38~  \mbox{ at 5 minutes} \\
	&L_{X,45}\!=\!11\,E_{\gamma,52}\quad\sigma_{log (L_{X}/E_{\gamma})}\!=\!0.51~  \mbox{ at 1 hour} \nonumber \\
	& L_{X,45}\!=\!0.42\,E_{\gamma,52}\quad \sigma_{log (L_{X}/E_{\gamma})}\!=\!0.64~ \mbox{ at 11 hours}\nonumber \\
	& L_{X,45}\!=\!0.15\,E_{\gamma,52}\quad \sigma_{log (L_{X}/E_{\gamma})}\!=\!0.71 \mbox{ at 24 hours \nonumber}
	\end{eqnarray}
	where $\sigma_{log (L_{X}/E_{\gamma})}$ is the 1$\sigma$ scatter (measured in log-log space)
	\footnote{We use here and elsewhere in the text the notation $Q_x=Q/10^x$ in c.g.s. units as well as base 10 logarithms.}.
	The correlation between $L_X$ at 11 hours and $E_{\gamma}$ has been investigated by different authors using different samples (see \citealt{Nysewander2009}, \citealt{Margutti2013}, and \citealt{Wygoda2016} for recent investigations). These studies find statistically significant correlations between $L_X$ and $E_{\gamma}$.
	The slope, normalization and scatter of the correlations discussed in these other studies are consistent with the one found in the BAT6 sample. Notice also, that since the measurements are done in a fixed observation time in the rest frame, $t_X$, (and that up to some constant $A$ of order unity that depends on the shape of the light-curve $E_X=AL_Xt_X$), equivalent relations exist between $E_X$ and $E_{\gamma}$ (where $E_X$ is the isotropic equivalent energy emitted in X-rays) and it follows that $\sigma_{log (E_{X}/E_{\gamma})}=\sigma_{log (L_{X}/E_{\gamma})}$. Furthermore, since in many cases $E_{\gamma}$ is calculated only within the relatively narrow BAT 15-350 keV energy range, the reported energies should be related to the full isotropic energies by a k-correction factor. Any spread in the k-correction is expected to only increase the observed scatter as compared to the underlying one since both $E_X$ and the full (k-corrected) $E_{\gamma}$ are governed by the shared total energy reservoir. Thus taking the spread of the observed ratio as the limiting value is a conservative choice.

	A measurement of $E_X,E_{\gamma}$ requires knowledge of the redshift, which is known for only a limited sample of bursts. However the ratio between the two energies, which is proportional to the observable ratio $F_{X,\rm peak} t_{X,\rm peak}/\Phi_{\gamma}$ (where $F_{X,\rm peak}$ is the peak of the X-ray flux measured at an observer time $t_{X,\rm peak}$ and $\Phi_{\gamma}$ is the measured $\gamma$-ray fluence during the prompt phase) is largely independent of the total energy and of the redshift. To illustrate this point we consider the standard afterglow model in which X-rays are produced by synchrotron emission from the forward external shock. The afterglow peaks when the jet starts being decelerated by the external medium $t_{X,\rm peak}\propto (1+z)E_{\rm kin}^{1/3}$ (see equation \ref{eq:tdec} below) where $E_{\rm kin}$ is the (isotropic equivalent) kinetic energy of the blast wave. At these early times, the emission is typically dominated by synchrotron in the fast cooling regime. Using the equation for the flux in this regime we have \citep{GS2002}: $F_{X,\rm peak} t_{X,\rm peak}\propto (1+z)^{(2+p)/4}E_{\rm kin}^{(2+p)/4} t_{X,\rm peak}^{(6-3p)/4}d_L^{-2}$ where $z$ is the redshift and $d_L$ is the luminosity distance. Assuming a roughly constant efficiency $E_{\rm \gamma}\propto E_{\rm kin}$. $E_{\gamma}$ is related to the prompt $\gamma$-ray fluence via $E_{\gamma}\propto \Phi_{\gamma} d_L^2/(1+z)$. Putting everything together the dependence on the energy cancels out and we are left with $F_{X,\rm peak} t_{X,\rm peak}/\Phi_{\gamma}\propto (1+z)^{(2-p)/2}$ (where $p$ is the power law slope of the electrons' energy distribution). $p$ is inferred from afterglow modelling to typically be in the range $2.1\leq p\leq 2.6$. Thus, we conclude that $F_{X,\rm peak} t_{X,\rm peak}/\Phi_{\gamma}$ should not depend strongly on the unknown distance to the GRB. Therefore, we can extend our sample of GRBs significantly, to include also bursts with no redshift determination. The advantage of this method is that it is largely independent of selection effects (and if any such effects exist they are unlikely to be shared with the redshift complete sample discussed above).

\begin{figure*}
		\centering
		\includegraphics[width = .37\textwidth]{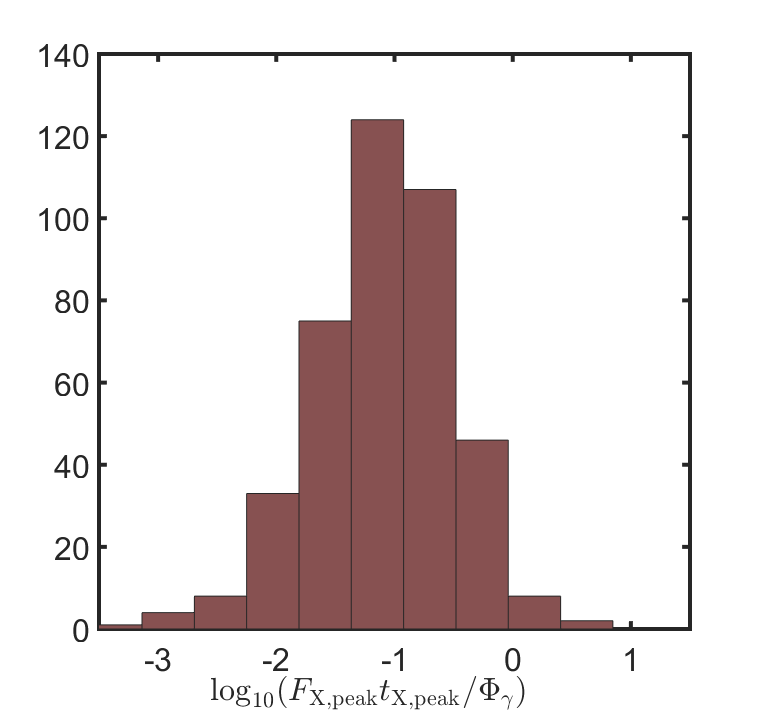}
		\caption{Histogram of early X-ray to prompt $\gamma$-ray fluence ratio ($F_{X,\rm peak} t_{X,\rm peak}/\Phi_{\gamma}$) for {\it Swift} BAT+XRT detected long bursts with a BAT $T_{90}$ duration larger than 3 sec and an XRT first detection before 200 s. The standard deviation in this ratio is $\sigma_{log (F_{X,\rm peak} t_{X,\rm peak}/\Phi_{\gamma})}=0.59$.}
		\label{fig:fluenceratio}
	\end{figure*}

	We use the {\it Swift} database \footnote{\url{https://swift.gsfc.nasa.gov/archive/grb_table/}} to collect all long GRBs detected in $\gamma$-rays by {\it Swift}-BAT (conservatively, we choose here a stringent requirement of $T_{90}>3$s, where $T_{90}$ is the prompt duration measured by BAT). Strikingly, we find that out of 765 bursts detected by BAT before July 2018, only 29 (representing a fraction of $\lesssim0.04$ of bursts) have no early detection with XRT. This alone strongly limits the intrinsic variability in $E_X/E_{\gamma}$. Selecting only bursts for which the BAT fluence, early XRT flux and observation time are automatically provided by the online {\it Swift} database, we are left with a sample of 456 bursts. In more than 408 ($90\%$) of these bursts the first X-ray detection is at $t<200$ s. For these bursts we find $\sigma_{log (F_{X,\rm peak} t_{X,\rm peak}/\Phi_{\gamma})}=0.59$ (see distribution in figure \ref{fig:fluenceratio}). Since in reality, this ratio depends on additional parameters other than the blast-wave energy, such as the redshift (see paragraph above), external density, bulk Lorentz factor, BAT k-correction etc. and since we have considered bursts with a range of peak times (up to 200 s) rather than one specific time, this standard deviation provides a strong {\it upper limit} on the true dispersion in $E_X/E_{\gamma}$ at 200 s. Note that any ``peculiarities" in some of these X-ray fluxes (such as early steep decline, X-ray flares etc.), assuming that they don't represent the vast majority of the data in this sample, would only cause a larger variance in the observed ratio $F_{X,\rm peak} t_{X,\rm peak}/\Phi_{\gamma}$. The similar observed scatter at these early times, compared to that measured at 5 minutes - 24 hours, suggests that the same underlying mechanisms are governing the observed trends in both cases and that thus, the X-ray luminosities at all these times are dominated by the standard forward shock afterglow.

	\section{The model}
	\label{sec:model}
	We assume a jet in which the kinetic energy per unit solid angle (in the central engine frame), $\epsilon$, and Lorentz factor, $\Gamma$ of the material may depend on the polar angle, $\theta$, from the jet's axis but is assumed to have azimuthal symmetry. We consider different initial distributions for $\epsilon, \Gamma$.
	\begin{enumerate}
		\item Constant up to some $\theta_0$ and then drops as a power law
		\begin{equation}
		\epsilon(\theta)=\frac{dE}{d\Omega}=\epsilon_0\left\{ \begin{array}{ll}1 & \theta<\theta_0\ ,\\
		\bigg(\frac{\theta}{\theta_0}\bigg)^{-\alpha} & \theta \geq \theta_0\ ,
		\end{array} \right.
		\end{equation}
		\begin{equation}
		\label{eq:GAPL}
		\Gamma(\theta)=1+(\Gamma_0-1)\left\{ \begin{array}{ll}1 & \theta<\theta_0\ ,\\
		\bigg(\frac{\theta}{\theta_0}\bigg)^{-\beta} & \theta \geq \theta_0\ ,
		\end{array} \right.
		\end{equation}
		Note that $\alpha>2$ is required in order for the jet's collimation-corrected energy to be dominated by the core (i.e., by material with $\theta<\theta_0$). The limiting case of $\alpha=2$ is known as a universal jet profile \citep{Rossi2002,Zhang2002}. $\alpha \geq 2$ is a necessary condition for the existence of jet breaks (which are seen in observations; \citealt{Rhoads1999,Sari1999,Panaitescu1999}), as well as for keeping the total energy and rate of GRBs within reasonable values. We therefore assume this situation applies in what follows.
		\item Gaussian, with a typical angular scale $\theta_0$
		\begin{equation}
		\epsilon(\theta)=\frac{dE}{d\Omega}=\epsilon_0e^{-(\theta/\theta_0)^2},
		\end{equation}
		\begin{equation}
		\label{eq:GAgaus}
		\Gamma(\theta)=1+(\Gamma_0-1)e^{-(\theta/\theta_0)^2}.
		\end{equation}
	\end{enumerate}
	\subsection{Prompt $\gamma$-rays}
	The (isotropic equivalent) kinetic energy is simply $E_{\rm k,iso}(\theta)=4\pi \epsilon(\theta)$. Assuming an efficiency $\eta_{\gamma}$ of conversion between the kinetic energy and the $\gamma$-rays, we get that the isotropic equivalent emitted energy in $\gamma$-rays is $E_{\gamma,\rm em}(\theta)=\eta_{\gamma}4\pi \epsilon(\theta)$. The observed $\gamma$-ray energy for an observer located at angle $\theta_{\rm obs}$ from the jet's axis (after correcting for cosmological redshift, i.e. defined in the ``central engine" frame) is then given by (see also \citealt{Salafia2015})
	\begin{equation}
	\label{eq:gammaenergy}
	E_{\gamma,\rm obs}(\theta_{\rm obs})=\eta_{\gamma}\int \frac{\epsilon(\theta)}{\Gamma(\theta)}\delta^3(\theta,\phi,\theta_{\rm obs})d\Omega
	\end{equation}
	where $\epsilon(\theta)$ is the energy per solid angle in the observer frame and $\delta$ is the Doppler factor
	\begin{equation}
	\delta(\theta,\phi,\theta_{\rm obs})=\frac{1}{\Gamma(\theta)(1-\beta(\theta) \cos \chi)}.
	\end{equation}
	In the last expression, $\beta(\theta)$ is the velocity corresponding to $\Gamma(\theta)$ and $\chi$ is the angle between the emitting material and the observer, and is given by
	\begin{equation}
	\cos \chi = \cos \theta_{\rm obs} \cos \theta + \sin \theta_{\rm obs} \sin \theta \cos \phi
	\end{equation}
	where by virtue of the azimuthal symmetry of the jet, we have assumed here without loss of generality that $\phi_{\rm obs}=0$.

	Due to compactness arguments, $\Gamma_0\gg\theta_0^{-1}$ \citep{Ruderman1975}. We define $q=|\theta_{\rm obs}- \theta_0|\Gamma(\theta_0)$. By performing the integration in Eq. \ref{eq:gammaenergy} we obtain the observed $\gamma$-ray energy \citep{Kasliwal2017,Granot2017,Ioka2018}:
	\begin{eqnarray}
	& E_{\gamma,\rm obs}(\theta_{\rm obs})= \\& \eta_{\gamma}4\pi \epsilon(\theta_0) \left\{ \begin{array}{ll}1 & \! \! |\theta_{\rm obs}|\ll |\theta_0|\\
	\max[\frac{\epsilon(\theta_{\rm obs})}{\epsilon(\theta_0)},q^{-4}] \! \! & \! \! |\theta_{\rm obs}\!-\! \theta_0|\ll \theta_0\\ \max[\frac{\epsilon(\theta_{\rm obs})}{\epsilon(\theta_0)},q^{-6}(\theta_0\Gamma_0)^2]\!\! &\! \! |\theta_{\rm obs}\!-\! \theta_0|\gg \theta_0
	\end{array} \right.  \nonumber
	\end{eqnarray}
	Generally, the emission may be dominated either by ``line of sight" emitters (i.e. that are radiating within an angle of up to  $1/\Gamma$ relative to the line of sight to the observer), in which case $E_{\gamma,\rm obs}(\theta_{\rm obs})=E_{\gamma,\rm em}(\theta)$ or by ``off line of sight" emitters (in which case the second case in each row applies).
	
	\begin{figure}
		\centering
		\includegraphics[width = .5\textwidth]{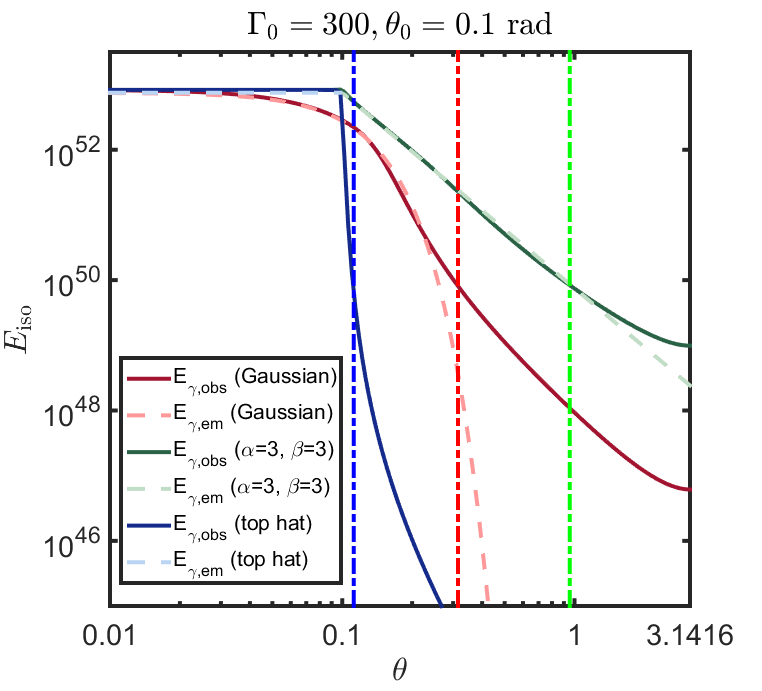}
		\caption{isotropic equivalent emitted (dashed) and observed (solid) $\gamma$-ray energies for different models of structured jets. We assume here $E_{\rm k,iso}=5\times 10^{53}$erg, $\eta_{\gamma}=0.15$, $\Gamma_0=300$, $\theta_0=0.1\mbox{ rad}$. The X-axis denotes $\theta_{\rm obs}$ for the $E_{\gamma,\rm obs}$ curves and $\theta$ for the $E_{\gamma,\rm em}$ curves. Also shown are the corresponding values of $\theta_{\rm max}$ for each case (dot dashed), which is the angle for which $E_{\gamma,\rm obs}(\theta_{\rm obs}=\theta_{\rm max})\equiv10^{-3}E_{\gamma,\rm obs}(\theta_{\rm obs}=0)$.}
		\label{fig:Egamma}
	\end{figure}

	Prompt observations reveal a wide range of observed (isotropic equivalent) $\gamma$-ray energies spanning at least three orders of magnitude (see discussion in \S \ref{sec:obs}). This implies that even if no variance exists between the parameters (e.g. total energy, Lorentz factor, etc.) of different bursts, we should expect to see bursts up to a typical observation angle $\theta_{\rm max}$ which is defined here as the angle for which $E_{\gamma,\rm obs}$ is reduced by three orders of magnitude compared to an observer looking down the jet's core: 
	\begin{equation}
	E_{\gamma,\rm obs}(\theta_{\rm obs}=\theta_{\rm max})\equiv10^{-3}E_{\gamma,\rm obs}(\theta_{\rm obs}=0).
	\end{equation}

	Examples of $E_{\gamma,\rm em}(\theta), E_{\gamma,\rm obs}(\theta_{\rm obs})$ for different models are shown in figure \ref{fig:Egamma}. As can be seen in this figure, unless the distribution of energy and Lorentz factor in the jet declines very steeply at high latitudes, the observed $\gamma$-rays are typically dominated by the line of sight material for a large range of observing angles.

	\subsection{X-ray afterglow}
	A jet propagating through a constant external medium moves initially at an approximately constant velocity. This lasts until the jet reaches the so called, deceleration radius, where it has swept up enough of the external medium to begin slowing down. Assuming we are observing the emitting material along the line of sight (i.e. from within an angle of $1/\Gamma$), this happens at an observer time:
	\begin{eqnarray}
	\label{eq:tdec}
	&\frac{t_{\rm dec}}{1\!+\!z}=\bigg(\frac{17 E_{\rm k,iso}}{64 \pi n m_p c^5}\bigg)^{1/3}\Gamma^{-8/3}  \nonumber\\ &  = 15 E_{\rm k,iso,53}^{1/3}n^{-1/3}\Gamma_{2.5}^{-8/3} \mbox{ s}
	\end{eqnarray}
	where $n$ is the particle density of the medium, $z$ is the cosmological redshift and where here elsewhere we use the notation $q_x\equiv q/10^x$ in cgs units. Introducing structure to jet's energy and Lorentz factor distributions, implies that different portions of the jet may decelerate at different times \footnote{Note that at these time-scales the jet can safely be assumed to be expanding radially without interaction between the different angular sections of the jet \citep{Salmonson2003,VanEerten2012}}. In particular for the PL model (and $\theta \gg \theta_0$) we have $t_{\rm dec}\propto \theta^{8\beta-\alpha \over 3}$. Since the bolometric luminosity of the jet increases as $t^3$ for $t<t_{\rm dec}$ the overall emission from the line of sight material at those times is decreased by a factor
	\begin{eqnarray}
	\label{eq:BolLumeff}
	&\frac{L(t<t_{\rm dec})}{L(t_{\rm dec})}=\bigg(\frac{t}{t_{\rm dec}}\bigg)^3=\\& 
	=\left\{ \begin{array}{ll}\!\!\!2.5\! \times \! 10^{3}E_{0,53}^{-1}n\Gamma_{0,2.5}^8 t_{\rm 200 s}^3 \bigg(\!\frac{\theta}{\theta_0}\!\bigg)^{\alpha-8\beta}\! \! (1\!+\!z)^{-3} & \mbox{PL}\\
	\!\!\!2.5\!\times\! 10^{3}E_{0,53}^{-1}n\Gamma_{0,2.5}^8 t_{\rm 200 s}^3 e^{-7(\theta/\theta_0)^2}(1\!+\!z)^{-3} & \mbox{Gauss.}
	\end{array} \right.\nonumber
	\end{eqnarray}
	where $E_0,\Gamma_0$ are the isotropic equivalent kinetic energy of the jets' core and the Lorentz factor of the core respectively and $t_{\rm 200 s}$ is the observer time measured in units of 200 s. At first glance, it seems that the terms in the bottom lines of Eq. \ref{eq:BolLumeff} are not smaller than unity. However, recall that for structured jets most bursts will be detected close to $\theta_{\rm max}$. Due to the extremely strong dependence of $L(t)/L(t_{\rm dec})$ on $\theta_{\rm max}$, this has a very significant effect on the bolometric efficiency. As an example for $\alpha=3,\beta=3,\Gamma_0=300,\theta_0=0.1\mbox{ rad}$, $\theta_{\rm max}=1\mbox{ rad}$ and the mean observation angle is $\langle \theta \rangle=0.65\mbox{ rad}$, implying that $L(t)/L(t_{\rm dec})= 2\times 10^{-14} t_{\rm 200 s}^3$.
	Alternatively, for a Gaussian jet with $\Gamma_0=300, \theta_0=0.1\mbox{ rad}$, $\theta_{\rm max}=0.31\mbox{ rad}$ we have $\langle \theta \rangle=0.2\mbox{ rad}$ implying that $L(t)/L(t_{\rm dec})=2\times 10^{-9} t_{\rm 200 s}^3$. Both situations result in a strong reduction of the bolometric luminosity at early times for observers at a viewing angle close to $\theta_{\rm max}$.
	
	\begin{figure}
		\centering
		\includegraphics[width = .5\textwidth]{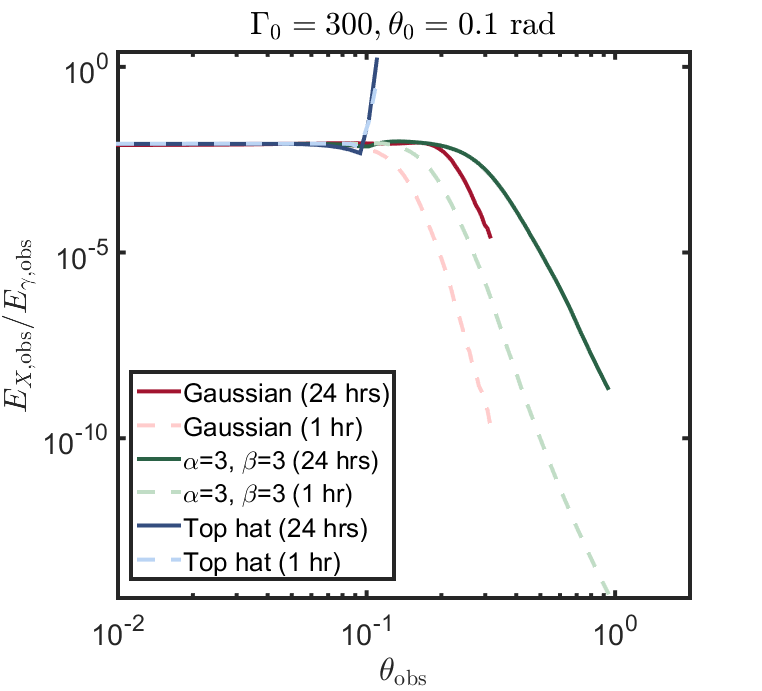}
		\caption{Ratio of observed X-ray afterglow to prompt $\gamma$-ray energies for different models of structured jets and different observation times. All curves are cut-off at $\theta_{\rm obs}=\theta_{\rm max}$, which is defined as the angle for which the observed gamma-rays is reduced by three orders of magnitude compared to an observer along the core of the jet.}
		\label{fig:EgammaEx}
	\end{figure}

	The ratio between the observed $E_{\gamma}$ and $E_X$ is depicted in figure \ref{fig:EgammaEx}. As is clear from the figure, even when limiting the observed angle to $\theta_{\rm obs}<\theta_{\rm max}$, the Gaussian and the PL model with $\alpha=\beta=3$ lead to a wide range of $E_X/E_{\gamma}$ ratios between different characteristic observers. Furthermore, the behaviour is more pronounced when X-rays are observed early on, since a larger portion of the jet has not yet decelerated at that stage.

	The implication is that {\it if $\gamma$-rays are seen up to a large angle from the jet's axis, then the Lorentz factor at that angle has to be large enough for the material in front of the observer to decelerate before the time of observation}.
	This roughly translates to:
	\begin{eqnarray}
	\label{eq:Gammathobs}
	&\Gamma(\theta_{\rm obs})\gtrsim \\& 
	\left\{ \begin{array}{ll} 110 \bigg(\frac{E_{0,53}}{n}\bigg)^{1\over 8} \bigg(\frac{1\!+\!z}{t_{\rm 200 s}}\bigg)^{3\over 8} \bigg(\frac{\theta_{\rm obs}}{\theta_0}\bigg)^{-{\alpha \over 8}} & \mbox{PL}\\
	110 \bigg(\frac{E_{0,53}}{n}\bigg)^{1\over 8} \bigg(\frac{1\!+\!z}{t_{\rm 200 s}}\bigg)^{3\over 8} e^{-\frac{1}{8}(\theta_{\rm obs}/\theta_0)^2} & \mbox{Gauss.}
	\end{array} \right.\nonumber
	\end{eqnarray}
	Given the small observed ratio of $E_X(t=200 \rm s)/E_{\gamma}$ discussed in \S \ref{sec:obs}, the majority of observed bursts must have decelerated by $t=200$ s. Thus, even if $\theta_{obs}$ is large compared to $\theta_0$, the Lorentz factor at those angles must still remain rather large, to comply with these observations.
	Considering the observation angle for which the energy drops by three orders of magnitude compared to the core $\theta_{\rm obs}=\theta_{\rm max}$, equation \ref{eq:Gammathobs} leads to $\Gamma(\theta_{\rm obs})\gtrsim 50$ independently of the specific energy profile.
	Given a typical Lorentz factor at the core of $\Gamma_0\approx 300$ \citep{Ghirlanda2018}, these results clearly show that {\it the Lorentz factor distribution must be much shallower than that of the energy}. Note that these limits depend very weakly on the kinetic energy at the core of the jet and on the external density. For long GRBs the external density can in some cases be dominated by a wind profile $\rho \propto A_* r^{-2}$. In the wind case, equation \ref{eq:Gammathobs} can be re-written as
	\begin{eqnarray}
	\label{eq:Gammathobswind}
	&\Gamma(\theta_{\rm obs})\gtrsim \\& 
	\left\{ \begin{array}{ll} 74 \bigg(\frac{E_{0,53}}{A_{*,-1}}\bigg)^{1\over 4} (\frac{1\!+\!z}{t_{\rm 200 s}}\bigg)^{1\over 4} \bigg(\frac{\theta_{\rm obs}}{\theta_0}\bigg)^{-{\alpha \over 4}} & \mbox{PL}\\
	74\bigg(\frac{E_{0,53}}{A_{*,-1}}\bigg)^{1\over 4} (\frac{1\!+\!z}{t_{\rm 200 s}}\bigg)^{1\over 4} e^{-\frac{1}{4}(\theta_{\rm obs}/\theta_0)^2} & \mbox{Gauss.}
	\end{array} \right.\nonumber
	\end{eqnarray}
	where we have taken here $A_*=0.1$ as typically inferred from afterglow observations \citep{GvdH2014}.
	Using the above equation with $\theta_{\rm obs}<\theta_{\rm max}$ implies $\Gamma(\theta_{\rm obs})\gtrsim 10$.
	
	Assuming angular distributions for the Lorentz factor as in equations \ref{eq:GAPL},\ref{eq:GAgaus} for the PL and Gaussian cases correspondingly, equation \ref{eq:Gammathobs} can be re-written as
	\begin{eqnarray}
	\label{eq:betalphalim}
	&\beta<\frac{1}{8}\alpha-\frac{\ln\bigg(0.36\bigg(\frac{E_{0,53}}{n}\bigg)^{1\over 8}\bigg(\frac{1\!+\!z}{t_{\rm 200 s}}\bigg)^{3\over 8} \Gamma_{0,2.5}^{-1} \bigg)}{\ln\big(\theta_{\rm obs}/\theta_0\bigg)} \ \ \mbox{ PL}  \\
	& \bigg(\frac{\theta_{\rm obs}}{\theta_0}\bigg)^2<-\frac{8}{7}\ln\bigg(0.36\bigg(\frac{E_{0,53}}{n}\bigg)^{1\over 8}\bigg(\frac{1\!+\!z}{t_{\rm 200 s}}\bigg)^{3\over 8} \Gamma_{0,2.5}^{-1} \bigg) \ \ \mbox{ Gauss} \nonumber
	\end{eqnarray}
	In what follows we revisit these limits with a more realistic and detailed approach. As we will show in \S \ref{sec:results}, the equation above indeed provides a good approximation for the limitations on the different models.
	In particular, the second line shows that for the Gaussian case, and $t_{\rm X}=200$ s, $\Gamma_0=300$, Gaussian profiles for the energy / Lorentz factor, are only consistent if $\theta_{\rm obs}\lesssim 1.2\theta_0$ (the exact limit is re-addressed in \S \ref{narrowgamma}).
	\section{Permitted structure models in light of $E_X/E_{\gamma}$ observations}
	\label{sec:results}
	In order to determine which structure models are allowed by observations we perform a Monte Carlo simulation, and compare the results for the combined $E_{\gamma},E_X$ distribution with the observed samples described in \S \ref{sec:obs}.
	As a conservative approach, we assume here that there is no intrinsic scatter in the isotropic equivalent core luminosity emitted in the $\gamma$-rays, $L_{\gamma,em}(\theta=0)$, and take the median value for bursts in the BAT6 sample for the component along the jets' core, $L_{\gamma,\rm em}(\theta=0)=2\times 10^{51}\mbox{ erg s}^{-1}$ in the central engine frame. This implies that the observed range of prompt energies is solely dictated by the structure of the jet. Any intrinsic scatter in this property will leave less room for structure variations, and will thus strengthen our conclusions below (we test this assumption below). For the same reason, we do not allow here for any variation in the peak $\gamma$-ray energy $E_p$ (and take a constant $E_p(\theta=0)=500$keV in the central engine frame, close to the median of the BAT6 sample) or the afterglow parameters: $E_{\rm k,iso}(\theta=0),\epsilon_e,\epsilon_B,n,p$ (taken here as $5\times 10^{53}\mbox{erg},0.1,10^{-4},1\mbox{cm}^{-3},2.2$ respectively, see \citealt{Nava2014,Santana2014,Beniamini2015,Zhang2015,BvdH2017}). 
	We also take a normal distribution for the prompt $\gamma$-rays' low energy spectral slope with a peak at $\alpha=-1.1$ and a standard deviation of $0.47$ dex. For redshifts, we take a log-normal distribution in $z$ with a peak at $z=1.8$ and a standard deviation of $0.3$ dex. The two latter distributions are fits to the observed bursts in the BAT6 sample and are also consistent with the general GRB population.
	Finally, the viewing angle is assumed to be isotropically distributed. For each model, we only consider bursts for which the $\gamma$-ray emission is above a limiting photon peak flux $P_{\rm lim}=2.6 \mbox{ photons s}^{-1}\mbox{cm}^{-2}$ in the 15-150keV range, as is the selection criteria in the BAT6 sample.

\begin{figure*}
		\centering
		\includegraphics[width = .39\textwidth]{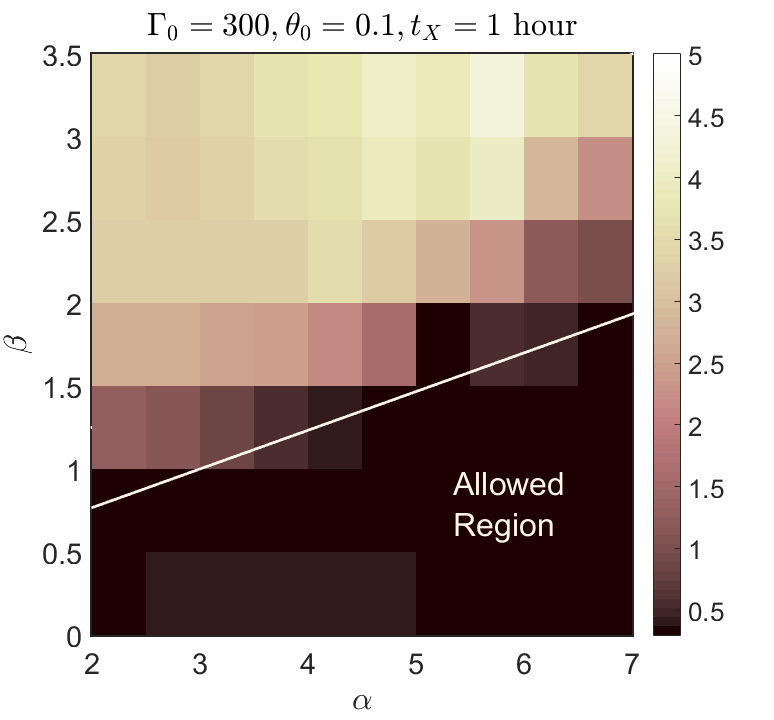}
		\includegraphics[width = .39\textwidth]{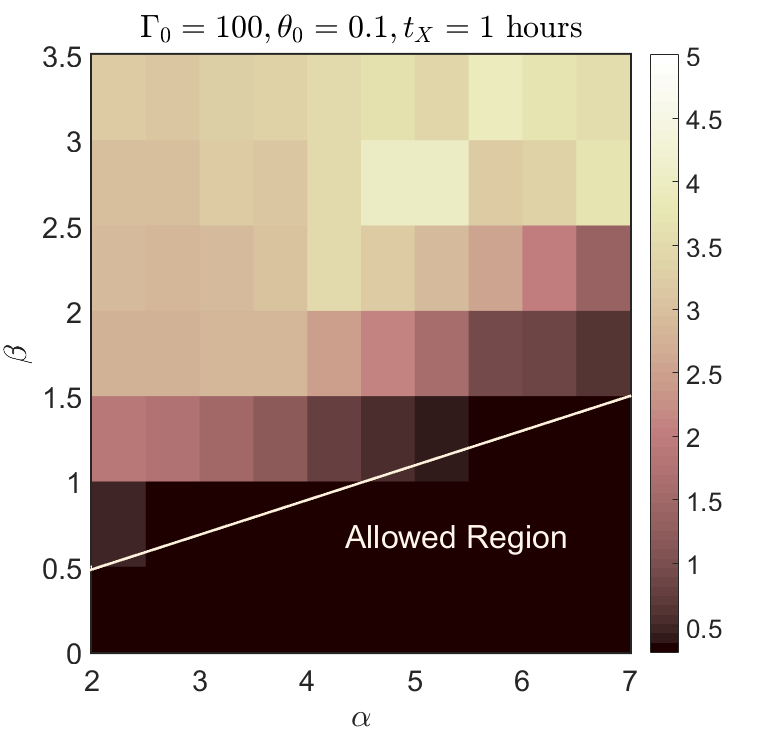}\\
		\includegraphics[width = .39\textwidth]{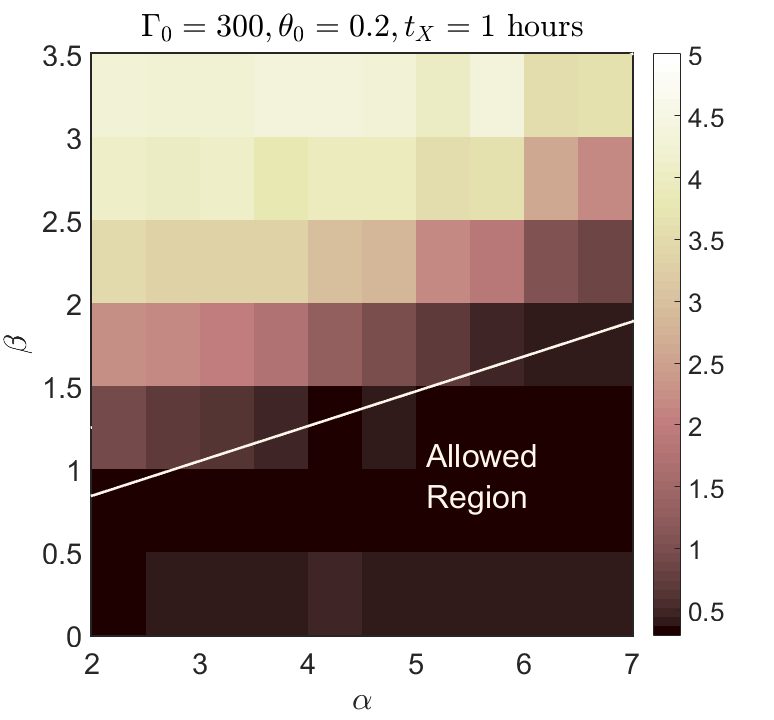}
		\includegraphics[width = .39\textwidth]{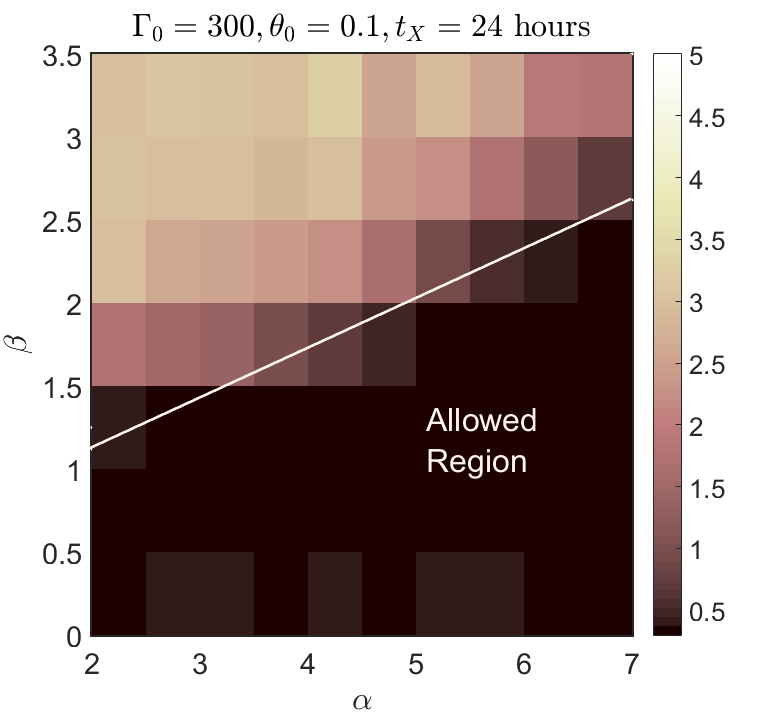}
		\caption{Lower limits on the width of the distribution of the ratio of X-ray afterglow to prompt $\gamma$-ray energies ($\sigma_{log (E_{X}/E_{\gamma})}$) for simulated bursts arising from different structure models and for which the $\gamma$-rays are within the observable range (see \S \ref{sec:results}). We assume here: $\Gamma_0=300,100\,\theta_0=0.1\mbox{ rad},0.2\,\rm{ }t_{\rm X}=1,24$ hours. The solid curves depict the observed scatter for the same $t_{\rm X}$ ($\sigma_{log (E_{X}/E_{\gamma})}\!=\!0.51,0.71$ for $1$ and $24$ hours respectively). Allowed structure models lie below these curves.}
		\label{fig:cumratio}
	\end{figure*}

\begin{figure*}
		\centering
		\includegraphics[width = .39\textwidth]{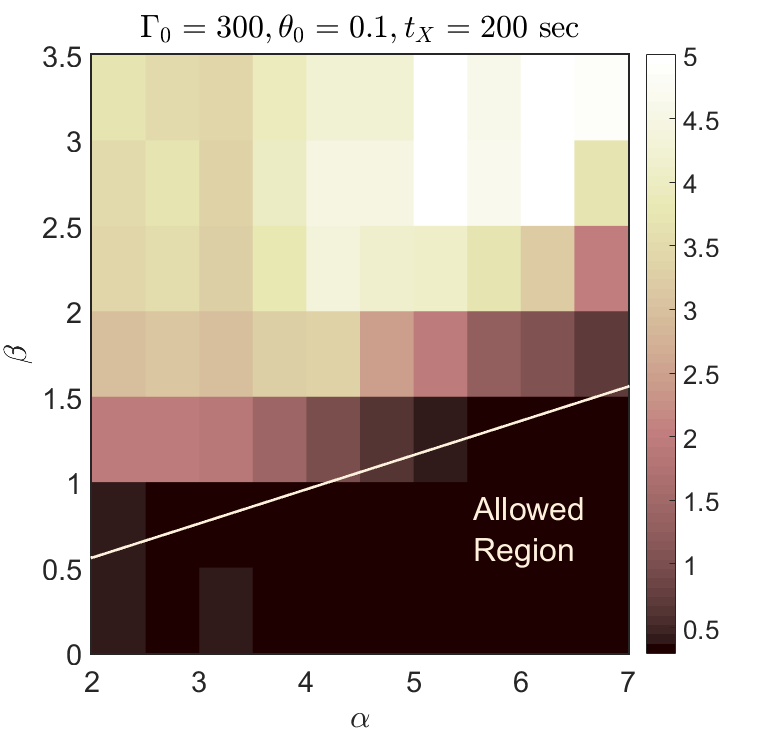}
		\caption{Same as figure \ref{fig:cumratio} for an observation time of $t_{\rm X}=200$ s. In this case, the limiting curve from the observed scatter becomes $\sigma_{log (E_{X}/E_{\gamma})}\!=\!0.59$ for $200$ s.}
		\label{fig:cumratiot200}
	\end{figure*}

	We draw $10^4$ random viewing angles and compute the observed prompt-afterglow energy distribution. The values of $\sigma_{log (E_{X}/E_{\gamma})}$ are shown in figures \ref{fig:cumratio},\ref{fig:cumratiot200} for different observation times and different PL models, exploring also different values for the X-ray observation time, jet's core opening angle, $\theta_0$ and bulk Lorentz factor, $\Gamma_0$.
	We compare the values of $\sigma_{log (E_{X}/E_{\gamma})}$ for each model with the observed standard deviations discussed in \S \ref{sec:obs}, and consider models in which the former is larger than the latter to be ruled out. Note that since the observed $E_{X}/E_{\gamma}$ ratio is approximately normally distributed, the standard error in the standard deviation is $\lesssim 0.5\sigma, \lesssim 0.25\sigma$ for the BAT6 and the {\it Swift} database samples respectively. Given the strong dependence of $\sigma_{log (E_{X}/E_{\gamma})}$ on the combination of $\alpha,\beta$, the requirement described above for comparing the simulations with observations is rather robust \footnote{Note that the shape of the simulated distributions are typically quite different compared to that of the observed one. Indeed, we are not attempting to reproduce the observed distribution using the simulated ones, but just provide a minimal test for when there are statistically significant deviations between the two. The comparison of the standard deviations is conservative, as it is a necessary but insufficient condition for the observed and simulated distributions to be compatible. Considering other moments of the distributions could thus only further constrain the allowed models.}.
	As can be seen in the figures, a significant population of PL models are ruled out by the data. In particular, for a typical $\theta_0=0.1\mbox{ rad},\Gamma_0=300$ and for $\alpha=2$, models with even a relatively shallow LF distribution: $\beta>0.55$ ($\beta>0.75$) are ruled out by the $200$ s (1 hour) data. For larger $\alpha$ the limiting value of $\beta$ increases roughly as $\beta \geq 0.2 \alpha +0.16$ ($\beta \geq 0.23 \alpha +0.29$). These limits are comparable to those found in equation \ref{eq:betalphalim} (but with a slightly steeper slope, since $\theta_{\rm obs}/\theta_0$ becomes smaller for larger values of $\alpha$).
	As expected, only very shallow Lorentz factor distributions are permissible by the data.

\begin{figure*}
		\centering
		\includegraphics[width = .39\textwidth]{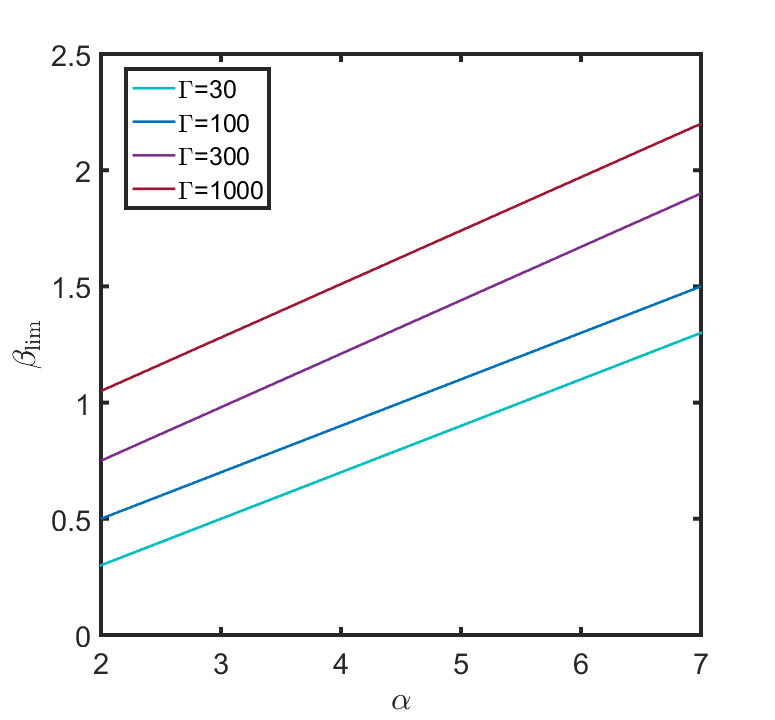}
		\caption{Dependence of the limiting curves (as shown in figure \ref{fig:cumratio} for $\Gamma_0=300,\theta_0=0.1\mbox{ rad}, t_{\rm X}=1$ hr, and above which models are ruled out by their scatter in $E_{X}/E_{\gamma}$) as a function of the bulk Lorentz factor at the jet's core, $\Gamma_0$.}
		\label{fig:Gamma0dep}
	\end{figure*}

	The dependence of the limiting curve on the value of the bulk Lorentz factor at the jet's core, $\Gamma_0$, is shown explicitly in figure \ref{fig:Gamma0dep}. Clearly, larger $\Gamma_0$ implies that a larger portion of the jet has already decelerated by the time of X-ray observations, and as a result, the X-ray luminosity better matches the isotropic equivalent energy and the scatter is reduced (hence, the limits on the $\alpha-\beta$ plane are less constraining). We note that the canonical value of $\Gamma_0=300$ adopted above is in accord with observations of deceleration peaks in long GRBs \citep{Ghirlanda2018}. For each of the cases explored in figure \ref{fig:cumratio}, we also test models where $\epsilon,\Gamma$ are Gaussian (see \S \ref{sec:model}). We find standard deviations of: $\sigma_{log (E_{X}/E_{\gamma})}\!=\!4.8,4.7,4,5.2,3.3$ for $[\Gamma_0=300,\theta_0=0.1\mbox{ rad}, t_X=200\mbox{ s}]$,$[\Gamma_0=300,\theta_0=0.1\mbox{ rad}, t_X=1\mbox{ hour}]$, $[\Gamma_0=100,\theta_0=0.1\mbox{ rad}, t_X=1\mbox{ hour}]$, $[\Gamma_0=300,\theta_0=0.2\mbox{ rad}, t_X=1\mbox{ hour}]$ and $[\Gamma_0=300,\theta_0=0.1\mbox{ rad}, t_X=24\mbox{ hour}]$ correspondingly, implying that in all these cases Gaussian models are strongly ruled out by observations. Note however, that a Gaussian energy structure with a flat Lorentz factor distribution is consistent with the observed structure (however, this model is limited by other considerations, as will be shown in \S \ref{sec:lightshape}). Additionally, we verified that in all above cases, top hat models are consistent with observations. Lastly, we note that out of the indirectly inferred GRB afterglow parameters, the most important quantity affecting the scatter in $E_{X}/E_{\gamma}$ is $n$, the external density (see equation \ref{eq:BolLumeff}). Here we have assumed a nominal $n=1\mbox{ cm}^{-3}$. Some studies have suggested that there may be a significant population of GRBs taking place at lower density environments \citep{Panaitescu2001,Cenko2010,VanderHorst2014,Laskar2016}. Note that our choice of the surrounding density is conservative, since lower values would lead to later deceleration times, and therefore would weaken $E_X$ and lead to an even larger scatter in $E_{X}/E_{\gamma}$. We have also tested the dependence of the results on the assumption of a typical structure. That is, we have verified that introducing scatter to $\theta_0,\alpha,\beta$ increases  the scatter in $E_{X}/E_{\gamma}$. In particular, assuming a log-normal distribution for $\theta_0$ with a median of $0.1$ and a standard deviation of $\sinh^{-1}(0.5)$ (corresponding to $\Delta \theta_0/\bar{\theta}_0\approx 0.5$), as well as Gaussian distributions for $\alpha,\beta$ with median values of $\bar{\alpha}=3,6$, $\bar{\beta}=1.5$ and with $\sigma_{\alpha}=\sigma_{\beta}=1$ results in an increase of $\sigma_{log (E_{X}/E_{\gamma})}$ by a large amount compared to the equivalent cases with no variance in $\theta_0,\alpha,\beta$.
	
\begin{figure*}
		\centering
		\includegraphics[width = .39\textwidth]{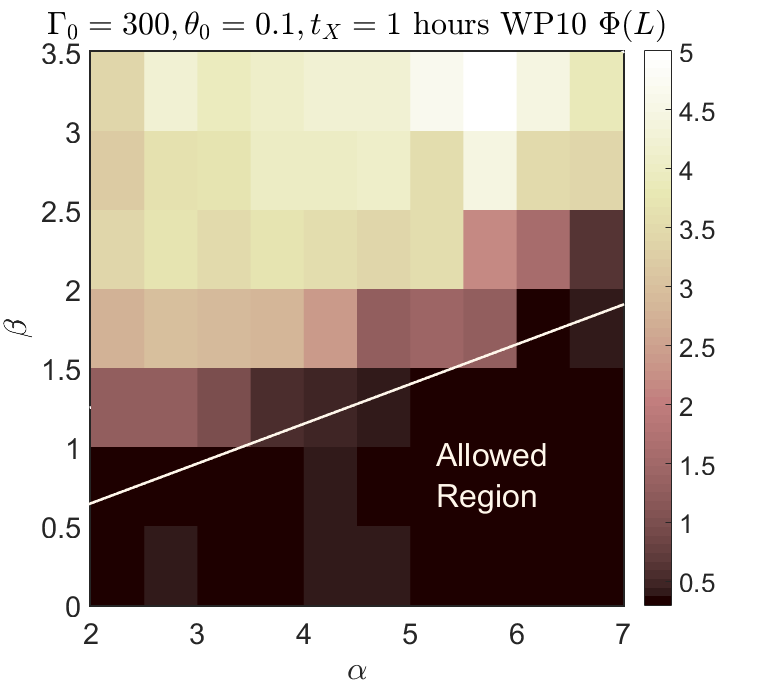}
		\caption{Same as figure \ref{fig:cumratio}, but with a core luminosity function taken according to literature values \citep{Wanderman2010}.}
		\label{fig:cumratioWP}
	\end{figure*}

	So far we've been conservatively assuming a constant intrinsic luminosity at the core. To test the importance of this assumption, we run another set of simulations where the luminosity function $\Phi(L_{\gamma})$ is taken according to values reported in the literature. As a test case, we use the luminosity function derived by \cite{Wanderman2010}. The resulting $E_{X}/E_{\gamma}$ scatter (for $\Gamma_0=300, \theta_0=0.1\mbox{ rad}, t_{\rm X}=1$ hour) can be seen in figure \ref{fig:cumratioWP}. For $\alpha=2$ one requires $\beta<0.65$, while for $\alpha=7$ this becomes $\beta<1.9$. Evidently, the results with this luminosity function are slightly more constraining (but overall quite similar) as compared with the equivalent case with $\theta_0=0.1\mbox{ rad},\Gamma_0=300,t_{\rm X}=1\mbox{ hour}$ reported above.

	\section{Structure constraints from the observed luminosity function}
	\label{sec:lumfunc}
	If GRBs typically have wide structures, it would imply that they are often observed far from their cores and detected at lower luminosities compared to their core emission. Indeed various authors have envisioned a scenario where a wide range of GRBs have the same core luminosity, and in which the observed luminosity function is simply the result of the observer's viewing angle \citep{Lipunov2001,Frail2001,Rossi2002,Zhang2002,Eichler2004,VanEerten2012,Pescalli2015}. Although this is an intriguing possibility, it is difficult to prove, as the observed luminosity function is a convolution of the true core luminosity distribution and the structure effect, and there are multiple combinations of both that can adequately describe the observed data.

\begin{figure}
		\centering
		\includegraphics[width = .39\textwidth]{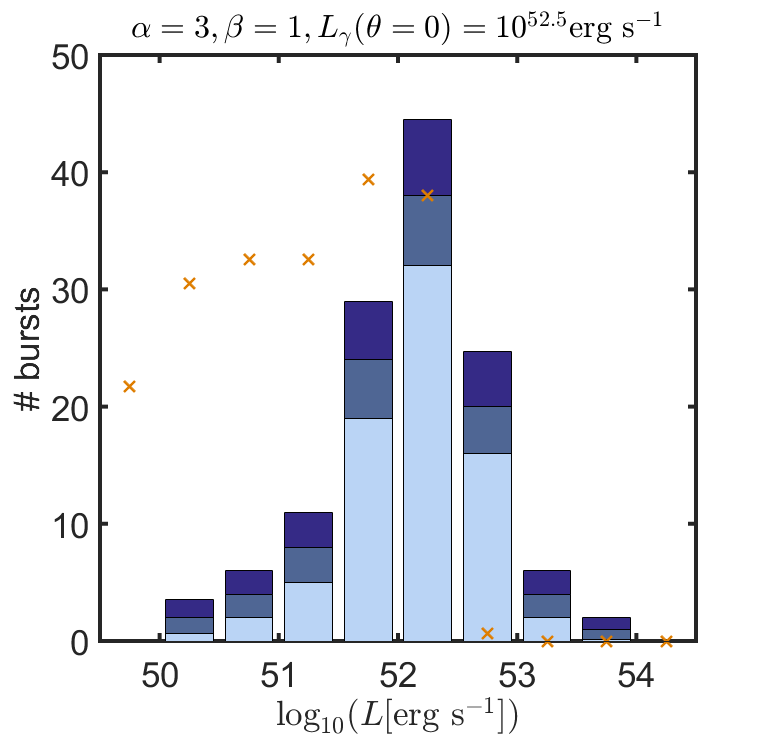}
		\caption{Crosses denote the number of observed bursts as a function of their $\gamma$-ray luminosity for a structure model $\alpha=3,\beta=1$ and assuming all bursts have the same luminosity along the jet's core $L_{\gamma}(\theta=0)=10^{52.5}\mbox{erg s}^{-1}$. Bars denote the number of observed bursts at the same luminosity (with statistical errors) in the sample of \citep{Wanderman2010}. The simulated sample is normalized such that the number equals the observed one at $L=L_*=10^{52.5}\mbox{erg s}^{-1}$.}
		\label{fig:Lumfunc}
	\end{figure}

	More conservatively, it is evident that the structure of GRBs should never overproduce lower luminosity bursts as compared with observations. The {\it observed} luminosity function \footnote{Not to be confused with the derived intrinsic luminosity function $\Phi(L)$ described elsewhere in the text.} peaks at $L_*\approx 3\times 10^{52} \mbox{erg s}^{-1}$ \citep{Wanderman2010}, and then falls sharply down until reaching the highest observed luminosity considered in the same study ($L_{\rm max}\approx 10^{54} \mbox{erg s}^{-1}$, see also figure \ref{fig:Lumfunc}). Therefore, under the assumption of a PL or Gaussian structure model, the distribution at $L_*$ must be dominated by events seen along the core of the jet that have a luminosity of $\approx L_*$. 
	The distribution below $L_*$ may be a mix of the intrinsic luminosity distribution at the jets' core and of bursts seen with $L\geq L_*$ that are viewed at angles beyond the core. Clearly, models in which the off-axis contribution of bursts at $L\approx L_*$ alone, is enough to over-prodcue bursts with $L<L_*$ are ruled out by observations.
	
	To test this, we simulate bursts with a single core luminosity that is equal to $L_*$, and with random viewing angles. We then select only those bursts that have $\gamma$-ray luminosities high enough to be detectable (see \S \ref{sec:results}). We then normalize the number of bursts with an observed $\gamma$-ray luminosity equal to $L_*$ to be the same as in the observed distribution reported by \cite{Wanderman2010} and compare the number of bursts with $L<L_*$ with the observed sample. An example of this for a structure model with $\alpha=3, \beta=1$ can be seen in figure \ref{fig:Lumfunc}, demonstrating that this particular model overproduces lower luminosity bursts and is therefore ruled out.

\begin{figure}
		\centering
		\includegraphics[width = .39\textwidth]{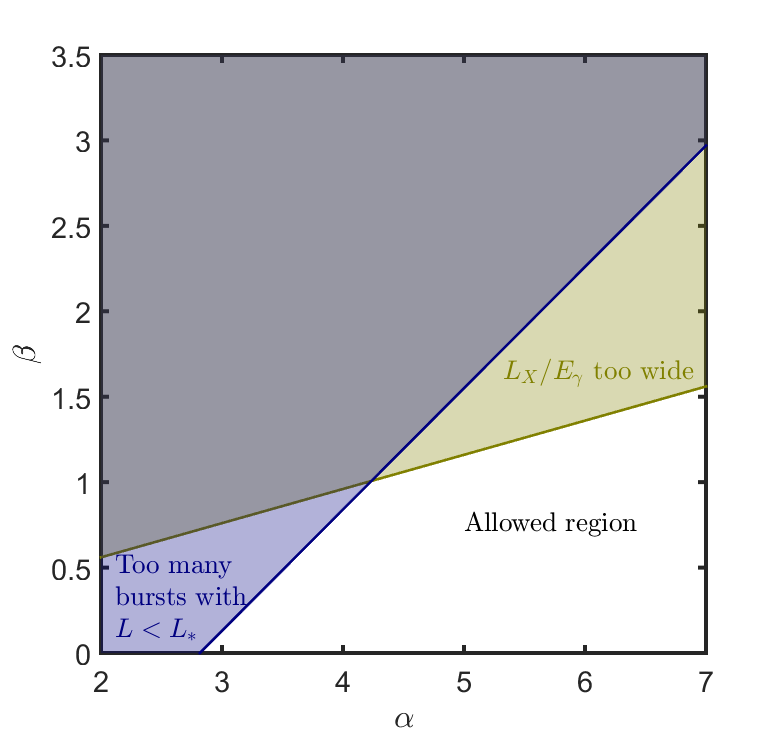}
		\caption{Ruled out structure models given that (a) models should not overproduce events with $L<L_*$ compared with observations (blue region) and (b) models should not produce a scatter in $E_X/E_{\gamma}$ that is larger than the observed one (Yellow).}
		\label{fig:Allowed}
	\end{figure}

	More generally, the allowed region in the $\alpha,\beta$ plane in order not to over-produce bursts with $L<L_*$ is shown in figure \ref{fig:Allowed}. The ruled out region (above the line $\beta \geq 0.71 \alpha -2$)
	is shown side by side with the region ruled-out by the $E_X/E_{\gamma}$ scatter observations (we consider here the default case with $\Gamma_0=300$, $\theta_0=0.1\mbox{ rad}$, $t_X=200$ s). For $\alpha\lesssim 4.3$ the luminosity function happens to be the more constraining consideration while for larger $\alpha$ values the trend is reversed. Both considerations, require steep structure models with a roughly constant Lorentz function as a function of latitude. Finally we note that Gaussian models for both the energy and Lorentz factor are ruled out by the luminosity function consideration, while a Gaussian model for the energy in which the Lorentz factor is constant, is consistent with the data. 
	
	\section{The shape of off-axis light-curves}
	\label{sec:lightshape}
	In \S \ref{sec:results} we have considered constraints on angular structure models arising from the scatter in $E_X/E_{\gamma}$, where $E_X$ was measured at a given observation time after the trigger. A more detailed treatment should consider the entire temporal evolution of the  afterglow emission in comparison with observed light-curves \citep[e.g.,][]{KG2003,GK2003,Salmonson2003,Lamb2017}. This is, in principle, a much more constraining consideration, but it is also less well defined from a statistical point of view. Nonetheless, as we show in this section, in jets with a significant drop in the kinetic energy outside of the jet core, $\theta_0$, the afterglows seen by observers at $\theta_{\rm obs} \gg \theta_0$ are different than those observed in GRBs (with the exception of GW170817). This is the case also if the Lorentz factor does not fall significantly outside of the core, and it implies that we do not detect long GRBs when the core points far away from us, which in turn strongly suggests that the $\gamma$-ray emission is inefficient at regions of the jet where the kinetic energy is much lower than that of the core .
	
	The structure of the jet has a strong effect on the evolution of the light curve when observed away from the core. The reason is that at any given time we observe a region at an angle $1/\Gamma$ with respect to the line-of-sight and as the blast wave driven into the circum-burst medium decelerates we see emission from angles that did not contribute at earlier times. Thus, if we see the jet at an angle where the energy is much lower than that of the core, then as $\Gamma$ decreases with time we see regions with increasing amounts of energy. 
	As a result the afterglow light curve decays more slowly (or even shows a rebrightening) compared to an afterglow seen by an observer that lies within the opening angle of the jet's core. The light curve of GW170817 is an example of such an afterglow. However, GW170817 is unique. In fact all GRB afterglows show a decay during the first few days after the burst\footnote{There are fluctuations, flattening and rebrightening episodes in some afterglows but when averaging the light curve over time scales of days, they always show a decay. On a shorter time scale of $10^{3}-10^{4}$s, some X-ray light curves show a relatively shallow decay rate of $t^{-0.2}-t^{-0.8}$ .}. \cite{Oates2012} and \cite{Racusin2016} examined a the average decay rate of a large sample of optical and X-ray light curves finding that all light curves in their samples decayed at least as fast as $t^{-1/2}$. 
	
\begin{figure*}
		\centering
		\includegraphics[width = .4\textwidth]{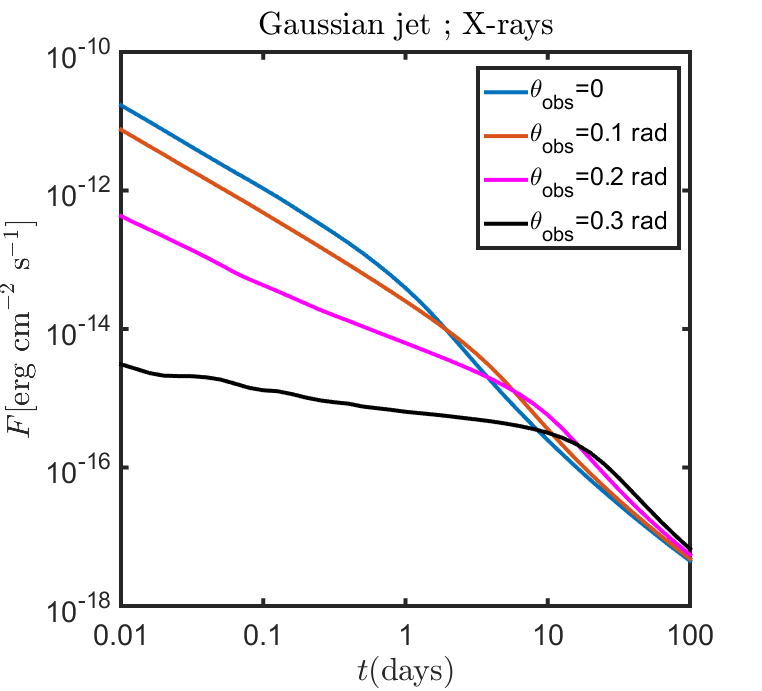}
		\includegraphics[width = .4\textwidth]{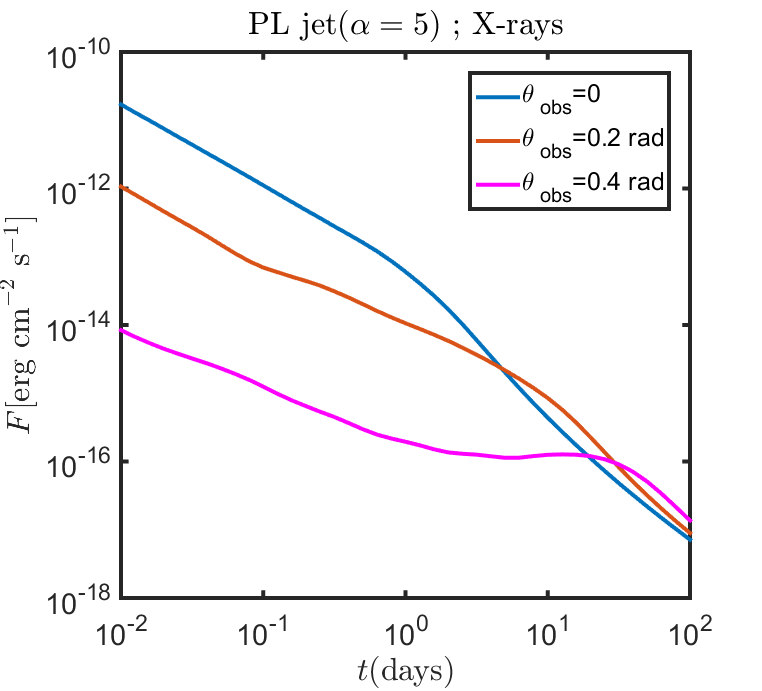}\\
		\includegraphics[width = .4\textwidth]{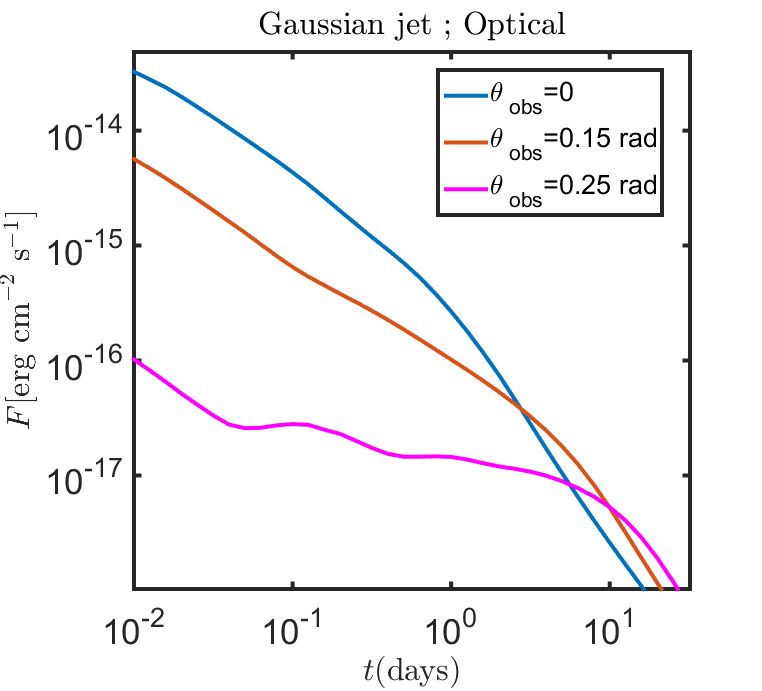}
		\includegraphics[width = .4\textwidth]{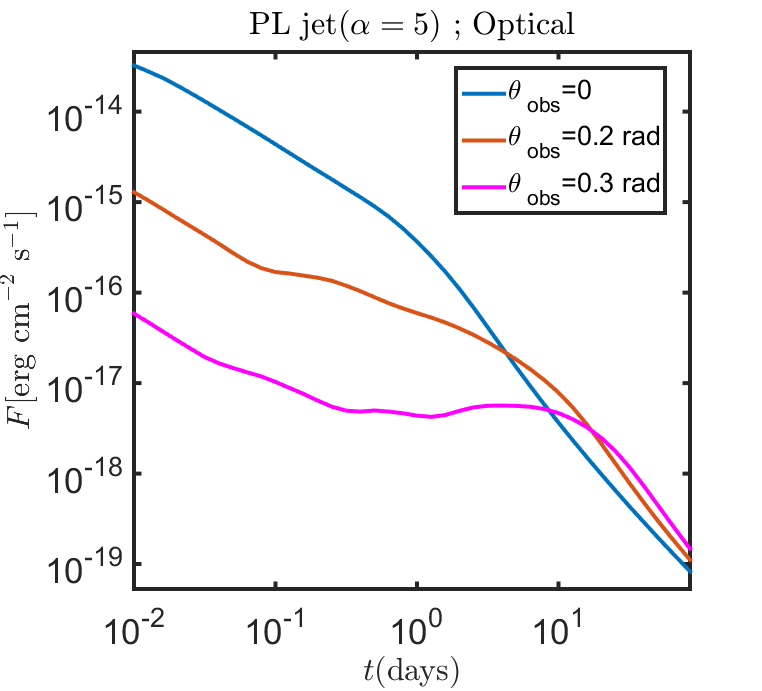}
		\caption{X-ray (top) and optical (bottom) light-curves of a GRB jet model with a Gaussian energy distribution (left) and a PL distribution with $\alpha=5$ (right). The Lorentz factor is constant at all angles in both models at $\Gamma(\theta)=300$. We also assume: $E_{\rm k,iso}=10^{52}$ erg, $n=1\mbox{ cm}^{-3}$, $p=2.2$, $\theta_0=0.1\mbox{ rad}$ and $z=1$. The range of viewing angles is chosen such that the prompt GRB remains observable in all cases.}
		\label{fig:lightcurve}
	\end{figure*}

	In previous sections we have shown that efficient gamma-ray emission is possible away from the core only if the the Lorentz factor remains very large at the emitting region. Here we examine a few representative afterglow light curves in cases that the Lorentz factor is large also away from the core. Figure \ref{fig:lightcurve} shows the X-ray and optical light curve shapes of two jets, one with a Gaussian and one with a power-law ($\alpha=5$) energy distributions. In both jets the Lorentz factor remains constant at all angles: $\Gamma(\theta)=\Gamma_0=300$. The angle of the jet core is $\theta_0=0.1\mbox{ rad}$ and its 
	energy is $E_{\rm k,iso}(\theta_0)=10^{52}$ erg. The external density is  $n=1\mbox{ cm}^{-3}$ and the electron power-law index is $p=2.2$. Additional parameters which affects the normalization only (but not the shape) are  $z=1$, $\epsilon_e=0.1$ and $\epsilon_B=0.01$. 
	
	When the observing angle is within the core of the jet, then the decay of the light curves of both jets are similar to each other and to many of the observed afterglows. However, when $\theta_{obs}>\theta_0$ the decay becomes shallower. For example at an angle of $\theta_{\rm obs}=0.3\mbox{ rad}$ the X-ray light curve of the Gaussian jet evolves roughly as $F_X\propto t^{-1/3}$ from very early time  up until $\sim 15$ days after the burst. Such an extended (in duration) shallow decline has to our knowledge not been observed in any GRB detected to date. The effect is even more pronounced in the optical, where even at an angle of $\theta_{\rm obs}=0.2\mbox{ rad}$ the average decay rate between $10^{-1}$ d and 3 d is $t^{-0.5}$ while for $\theta_{\rm obs}=0.25\mbox{ rad}$ it is $t^{-0.3}$. The fact that such afterglows if exist, are extremely rare, implies that if long GRB jet structure is Gaussian, then we do not detect gamma-ray emission from these jets at $\theta_{obs} \gtrsim 2\theta_0$ (and hence are not triggered to see their afterglows). This result applies even if only the energy density falls with the angle as a Gaussian while Lorentz factor remains constant.
	
	A similar behavior is seen in the Power-law jet. At observing angles $\theta_{\rm obs}>0.4\mbox{ rad}$ for X-rays and $\theta_{\rm obs}=0.3\mbox{ rad}$ for the optical, the light curves show a flattening and then a clear rise on time scales of 1-10 days. Such afterglows are again very rare, if they exist at all, implying that also in a power-law jet with $\alpha=5$ we do not detect $\gamma$-ray emission from these jets at $\theta_{obs} \gtrsim 3 \theta_0$.

	\section{An alternative possibility - restrictive gamma-ray region}
	\label{narrowgamma}
	
	\begin{figure}
		\centering
		\includegraphics[width = .37\textwidth]{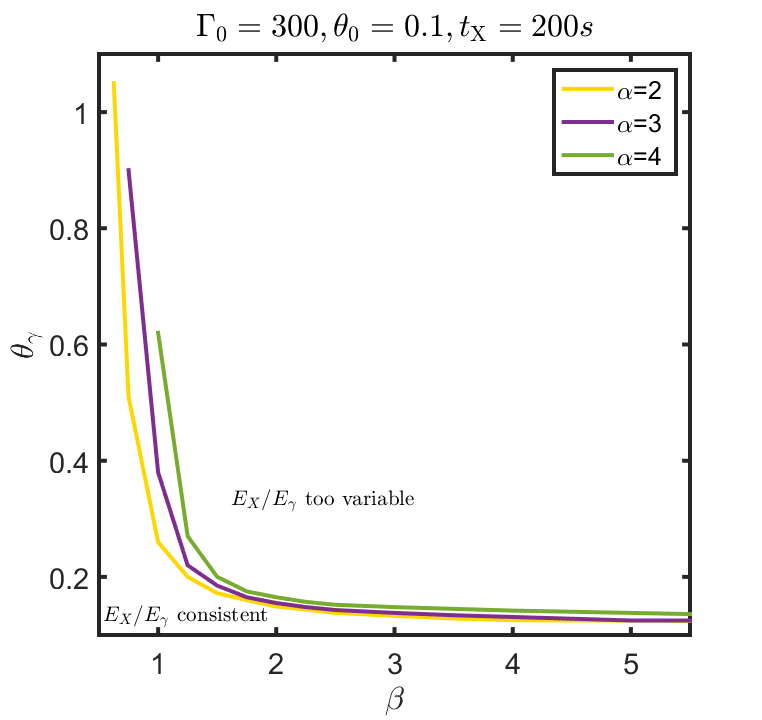}
		\includegraphics[width = .37\textwidth]{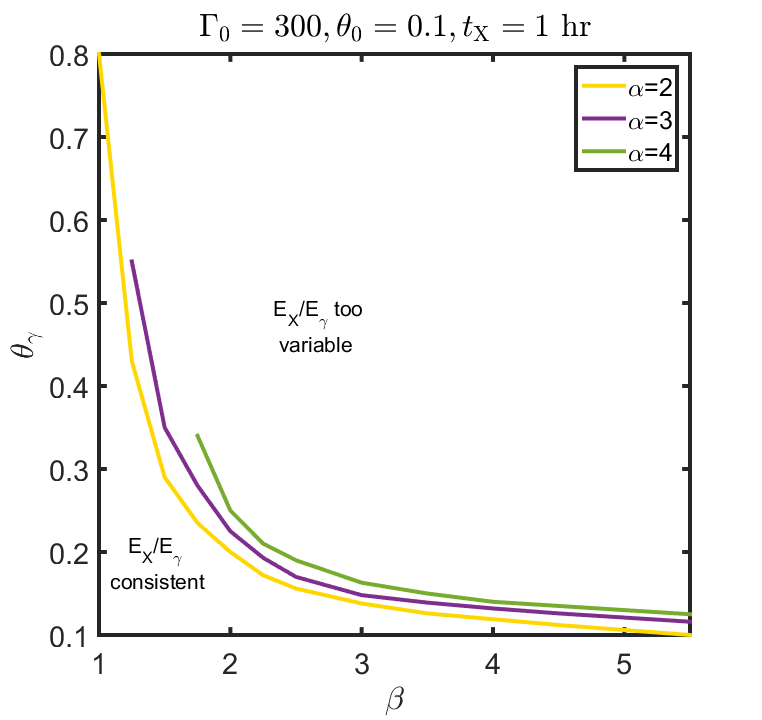}
		\caption{Maximum opening angle of $\gamma$-ray production ($\theta_{\gamma}$) required in order to keep the $E_X/E_{\gamma}$ scatter consistent with observations. Results are shown for PL models with either $t_{\rm X}=200$ s (top) or $t_{\rm X}=1$ hr (bottom), $\Gamma_0=300,\theta_0=0.1 \mbox{ rad}$ and varying values of $\alpha,\beta$.}
		\label{fig:thetagamma}
		\vspace{-0.2cm}
	\end{figure}

	In the previous sections we have shown that models with significant amounts of $\gamma$-ray emission beyond the jet's core are strongly constrained by observations. First, if $\gamma$-rays are produced efficiently far from the jet's core then the Lorentz factor cannot decrease very rapidly with angle, in order to avoid the material far from the jets' axis decelerating too slowly. Second, such jets observed off-axis would overproduce the low end of the luminosity function as compared with observations. Finally, we have shown, that even in the most extreme case, where the Lorentz factor remains constant across the jet, while the energy drops significantly, the typical shape of the light-curves associated with such bursts is very peculiar, and does not match any (non GW detected) GRB observed to date. At face value, these considerations seem to rule out almost any kind of structure that is not very close to a `top hat'.

	A possible way to get around these constraints is to assume that (efficient) $\gamma$-ray production is only confined to a narrow region around the jets' axis. Here we examine what is the largest opening angle of $\gamma$-ray production that is consistent with the observed scatter in $E_X/E_{\gamma}$ (we have verified that these limits are also consistent with the observed luminosity function and the shape of afterglow light-curves). We assume the same models for $\epsilon(\theta),\Gamma(\theta)$ as described in \S \ref{sec:model} (as well as $\Gamma_0=300, \theta_0=0.1 \mbox{ rad}, t_{\rm X}=200$ s or $t_{\rm X}=1$ hr), but we impose a cut-off on the $\gamma$-ray production region at a latitude of $\theta_{\gamma}$ such that $E_{\gamma,\rm em,new}(\theta)=E_{\gamma,\rm em}(\theta) \Theta(\theta_{\gamma}-\theta)$. Figure \ref{fig:thetagamma} depicts the largest allowed value for $\theta_{\gamma}$ for a PL model with $\alpha=2,3,4$ and varying values of $\beta$. As expected, larger values of $\beta$ correspond to a stricter (lower) cut-off on the $\gamma$-ray emission angle $\theta_{\gamma}$. In particular for $\alpha=\beta=3$, $\theta_{\gamma}=0.14\mbox{ rad}$ for $t_{\rm X}=200$ s ($\theta_{\gamma}=0.15\mbox{ rad}$ for $t_{\rm X}=1$ hr). As a comparison for the Gaussian model we find $\theta_{\gamma}=0.11\mbox{ rad}$ for $t_{\rm X}=200$ s ($\theta_{\gamma}=0.13\mbox{ rad}$ for $t_{\rm X}=1$ hr). These values are similar to the limits described in equation \ref{eq:betalphalim}.
	
	\vspace{-0.2cm}	
	\section{Conclusions}
	\label{conclusions}
	In this paper we examine various constraints that observations pose on the structure of GRB jets and the regions in these jets from which $\gamma$-rays are emitted efficiently.
	First, the ratio between the isotropic equivalent early X-ray afterglow and prompt $\gamma$-ray energy ($E_{X}/E_{\gamma}$) varies only moderately from burst to burst. We have shown here that a large set of GRB structure models predict a spread in this ratio that is significantly wider than the observed one, and can therefore be ruled out by this consideration. In particular the Lorentz factor of the emitting material cannot drop too fast (as a function of latitude) since it dictates a deceleration time that is too long and an extremely week output in X-rays during the first minutes to hours of the burst, in contrast with observations. 
	Specifically, in any region that radiates detectable $\gamma$-ray emission $\Gamma(\theta_{\rm obs})\gtrsim 50$. 
	Therefore, a Gaussian profile for the jet's energy is ruled out, unless the Lorentz factor profile is almost completely flat. Power-law profiles for the energy and Lorentz factor above $\theta_0$ (where $\epsilon\propto \theta^{-\alpha}$, $\Gamma\propto \theta^{-\beta}$) are ruled out above the critical line: $\beta \geq\beta_{E_X/E_{\gamma}}\equiv 0.2 \alpha +0.16$.

	Second, we consider which structure models are consistent with the observed luminosity function. Structures that lead to an over production of bursts with a luminosity below the peak of the observed luminosity function, are rejected by observations. This again rules out models with steep energy profiles and shallow Lorentz factor distributions, i.e. a Gaussian Lorentz factor profile and power-law  models above the line $\beta \geq \beta_{L-dist}\equiv 0.71 \alpha -2$.
	
	Finally, we show that even energy profiles with completely constant Lorentz factors at all angles of the jet, when their afterglows are observed at angles where the energy is much lower than that of the core (e.g., an angle of $\theta_{obs} \gtrsim 2-3\,\theta_0$ for a Gaussian jet and power-law jet with $\alpha=5$), show distinct light-curve shapes, which are evolving very shallowly for extended periods of time, or even exhibiting a shallow phase days after the trigger. Since such afterglows have not been observed, the existence of such GRB jets is highly questioned. 
	
	All of this suggests that most jet structures, such as Gaussian and power-law jets, are in strong tension with observations unless we cannot detect $\gamma$-rays at angles that are larger than about twice the core opening angle. This implies that if GRB jets have a structure with an energetic core and wings where the energy drops continuously, then efficient $\gamma$-ray emission is restricted only to a small area in and near the core of the jet. 
	Note, that this picture is consistent with the prompt $\gamma$-rays observed beyond the jet's core being produced by shock breakout from a mildly relativistic cocoon \cite{Nakar2012}. Cocoons produced by the jet propagation may result in significant amount of energy beyond the jets' cores, but they can be very inefficient in producing early $\gamma$-ray emission. 
	
	We have focused in this work chiefly on long GRBs. This is because the X-ray and optical data for short GRBs is much more sparse. It remains to be explored whether this is simply a selection effect since short GRBs are less energetic or alternatively, this is due to an intrinsic difference between short and long GRB structures. Nonetheless, there are some indications, that the scatter in $E_X/E_{\gamma}$ in short GRBs as measured at $1-24$ hr may be comparable to that in long GRBs \citep{Nakar2007,Berger2014}. In particular, this picture is consistent with the prompt $\gamma$-rays in GRB 170817 being produced away from the core, possibly by shock breakout from the mildly relativistic cocoon. 
	
	These results may also be used to inform understanding of future VIRGO / LIGO detected GRBs. clearly GRBs associated with GW detections could allow us to observe orders of magnitude fainter signals coming from GRBs, which would otherwise be undetected for cosmological events \citep[see also][]{Beniamini2018,Beniamini2018B} and quite literally, probe GRBs from a new angle. Once a large sample of GW detected GRBs is collected, the angular structure of the energy and Lorentz factor of GRBs may become directly testable by observations.
	\vspace{-0.4cm}
	\section*{Acknowledgements}	
	\vspace{-0.1cm}
	We thank Omer Bromberg, Pawan Kumar and Dimitrios Giannios for helpful discussions.


\bsp	
\label{lastpage}

\begin{thebibliography}{53}
		\expandafter\ifx\csname natexlab\endcsname\relax\def\natexlab#1{#1}\fi
		
		\bibitem[{Abbott {et~al}\mbox{.}(2017)Abbott {et~al.}}]{GW170817}
		Abbott B.~P., {et~al.}, 2017, Phys. Rev. Lett., 119, 161101
		
		\bibitem[{{Beniamini} {et~al}\mbox{.}(2018){Beniamini}, {Giannios}, {Younes},
			{van der Horst}, \& {Kouveliotou}}]{Beniamini2018}
		{Beniamini} P., {Giannios} D., {Younes} G., {van der Horst} A.~J.,
		{Kouveliotou} C., 2018, \mnras, 476, 5621
		
		\bibitem[{{Beniamini} {et~al}\mbox{.}(2015){Beniamini}, {Nava}, {Duran}, \&
			{Piran}}]{Beniamini2015}
		{Beniamini} P., {Nava} L., {Duran} R.~B., {Piran} T., 2015, \mnras, 454, 1073
		
		\bibitem[{{Beniamini}, {Nava} \& {Piran}(2016){Beniamini}, {Nava}, \&
			{Piran}}]{Beniamini2016}
		{Beniamini} P., {Nava} L., {Piran} T., 2016, \mnras, 461, 51
		
		\bibitem[{{Beniamini} {et~al}\mbox{.}(2019){Beniamini}, {Petropoulou}, {Barniol
				Duran}, \& {Giannios}}]{Beniamini2018B}
		{Beniamini} P., {Petropoulou} M., {Barniol Duran} R., {Giannios} D., 2019,
		\mnras, 483, 840
		
		\bibitem[{{Beniamini} \& {van der Horst}(2017)}]{BvdH2017}
		{Beniamini} P., {van der Horst} A.~J., 2017, \mnras, 472, 3161
		
		\bibitem[{{Berger}(2014)}]{Berger2014}
		{Berger} E., 2014, \araa, 52, 43
		
		\bibitem[{{Cenko} {et~al}\mbox{.}(2010){Cenko}, {Frail}, {Harrison},
			{Kulkarni}, {Nakar}, {Chandra}, {Butler}, {Fox}, {Gal-Yam}, {Kasliwal},
			{Kelemen}, {Moon}, {Ofek}, {Price}, {Rau}, {Soderberg}, {Teplitz}, {Werner},
			{Bock}, {Bloom}, {Starr}, {Filippenko}, {Chevalier}, {Gehrels}, {Nousek}, \&
			{Piran}}]{Cenko2010}
		{Cenko} S.~B. {et~al.}, 2010, \apj, 711, 641
		
		\bibitem[{{D'Avanzo} {et~al}\mbox{.}(2012){D'Avanzo}, {Salvaterra},
			{Sbarufatti}, {Nava}, {Melandri}, {Bernardini}, {Campana}, {Covino},
			{Fugazza}, {Ghirlanda}, {Ghisellini}, {La Parola}, {Perri}, {Vergani}, \&
			{Tagliaferri}}]{D'Avanzo2012}
		{D'Avanzo} P. {et~al.}, 2012, \mnras, 425, 506
		
		\bibitem[{{Eichler} \& {Levinson}(2004)}]{Eichler2004}
		{Eichler} D., {Levinson} A., 2004, \apjl, 614, L13
		
		\bibitem[{{Frail} {et~al}\mbox{.}(2001){Frail}, {Kulkarni}, {Sari},
			{Djorgovski}, {Bloom}, {Galama}, {Reichart}, {Berger}, {Harrison}, {Price},
			{Yost}, {Diercks}, {Goodrich}, \& {Chaffee}}]{Frail2001}
		{Frail} D.~A. {et~al.}, 2001, \apjl, 562, L55
		
		\bibitem[{{Ghirlanda} {et~al}\mbox{.}(2018){Ghirlanda}, {Nappo}, {Ghisellini},
			{Melandri}, {Marcarini}, {Nava}, {Salafia}, {Campana}, \&
			{Salvaterra}}]{Ghirlanda2018}
		{Ghirlanda} G. {et~al.}, 2018, \aap, 609, A112
		
		\bibitem[{{Granot} {et~al}\mbox{.}(2018){Granot}, {Gill}, {Guetta}, \& {De
				Colle}}]{Granot2017}
		{Granot} J., {Gill} R., {Guetta} D., {De Colle} F., 2018, \mnras, 481, 1597
		
		\bibitem[{{Granot} \& {Kumar}(2003)}]{GK2003}
		{Granot} J., {Kumar} P., 2003, \apj, 591, 1086
		
		\bibitem[{{Granot} \& {Sari}(2002)}]{GS2002}
		{Granot} J., {Sari} R., 2002, \apj, 568, 820
		
		\bibitem[{{Granot} \& {van der Horst}(2014)}]{GvdH2014}
		{Granot} J., {van der Horst} A.~J., 2014, \pasa, 31, e008
		
		\bibitem[{{Hallinan} {et~al}\mbox{.}(2017){Hallinan}, {Corsi}, {Mooley},
			{Hotokezaka}, {Nakar}, {Kasliwal}, {Kaplan}, {Frail}, {Myers}, {Murphy},
			{De}, {Dobie}, {Allison}, {Bannister}, {Bhalerao}, {Chandra}, {Clarke},
			{Giacintucci}, {Ho}, {Horesh}, {Kassim}, {Kulkarni}, {Lenc}, {Lockman},
			{Lynch}, {Nichols}, {Nissanke}, {Palliyaguru}, {Peters}, {Piran}, {Rana},
			{Sadler}, \& {Singer}}]{Hallinan2017}
		{Hallinan} G. {et~al.}, 2017, Science, 358, 1579
		
		\bibitem[{{Ioka} \& {Nakamura}(2018)}]{Ioka2018}
		{Ioka} K., {Nakamura} T., 2018, Progress of Theoretical and Experimental
		Physics, 2018, 043E02
		
		\bibitem[{{Kasliwal} {et~al}\mbox{.}(2017){Kasliwal}, {Nakar}, {Singer},
			{Kaplan}, {Cook}, {Van Sistine}, {Lau}, {Fremling}, {Gottlieb}, {Jencson},
			{Adams}, {Feindt}, {Hotokezaka}, {Ghosh}, {Perley}, {Yu}, {Piran}, {Allison},
			{Anupama}, {Balasubramanian}, {Bannister}, {Bally}, {Barnes}, {Barway},
			{Bellm}, {Bhalerao}, {Bhattacharya}, {Blagorodnova}, {Bloom}, {Brady},
			{Cannella}, {Chatterjee}, {Cenko}, {Cobb}, {Copperwheat}, {Corsi}, {De},
			{Dobie}, {Emery}, {Evans}, {Fox}, {Frail}, {Frohmaier}, {Goobar}, {Hallinan},
			{Harrison}, {Helou}, {Hinderer}, {Ho}, {Horesh}, {Ip}, {Itoh}, {Kasen},
			{Kim}, {Kuin}, {Kupfer}, {Lynch}, {Madsen}, {Mazzali}, {Miller}, {Mooley},
			{Murphy}, {Ngeow}, {Nichols}, {Nissanke}, {Nugent}, {Ofek}, {Qi}, {Quimby},
			{Rosswog}, {Rusu}, {Sadler}, {Schmidt}, {Sollerman}, {Steele}, {Williamson},
			{Xu}, {Yan}, {Yatsu}, {Zhang}, \& {Zhao}}]{Kasliwal2017}
		{Kasliwal} M.~M. {et~al.}, 2017, Science, 358, 1559
		
		\bibitem[{{Kumar} \& {Granot}(2003)}]{KG2003}
		{Kumar} P., {Granot} J., 2003, \apj, 591, 1075
		
		\bibitem[{{Lamb} \& {Kobayashi}(2017)}]{Lamb2017}
		{Lamb} G.~P., {Kobayashi} S., 2017, \mnras, 472, 4953
		
		\bibitem[{{Laskar} {et~al}\mbox{.}(2016){Laskar}, {Alexander}, {Berger},
			{Fong}, {Margutti}, {Shivvers}, {Williams}, {Kopa{\v c}}, {Kobayashi},
			{Mundell}, {Gomboc}, {Zheng}, {Menten}, {Graham}, \&
			{Filippenko}}]{Laskar2016}
		{Laskar} T. {et~al.}, 2016, \apj, 833, 88
		
		\bibitem[{{Lipunov}, {Postnov} \& {Prokhorov}(2001){Lipunov}, {Postnov}, \&
			{Prokhorov}}]{Lipunov2001}
		{Lipunov} V.~M., {Postnov} K.~A., {Prokhorov} M.~E., 2001, Astronomy Reports,
		45, 236
		
		\bibitem[{{Margutti} {et~al}\mbox{.}(2017){Margutti}, {Berger}, {Fong},
			{Guidorzi}, {Alexander}, {Metzger}, {Blanchard}, {Cowperthwaite}, {Chornock},
			{Eftekhari}, {Nicholl}, {Villar}, {Williams}, {Annis}, {Brown}, {Chen},
			{Doctor}, {Frieman}, {Holz}, {Sako}, \& {Soares-Santos}}]{Margutti2017}
		{Margutti} R. {et~al.}, 2017, \apjl, 848, L20
		
		\bibitem[{{Margutti} {et~al}\mbox{.}(2013){Margutti}, {Zaninoni}, {Bernardini},
			{Chincarini}, {Pasotti}, {Guidorzi}, {Angelini}, {Burrows}, {Capalbi},
			{Evans}, {Gehrels}, {Kennea}, {Mangano}, {Moretti}, {Nousek}, {Osborne},
			{Page}, {Perri}, {Racusin}, {Romano}, {Sbarufatti}, {Stafford}, \&
			{Stamatikos}}]{Margutti2013}
		{Margutti} R. {et~al.}, 2013, \mnras, 428, 729
		
		\bibitem[{{Mooley} {et~al}\mbox{.}(2018{\natexlab{a}}){Mooley}, {Deller},
			{Gottlieb}, {Nakar}, {Hallinan}, {Bourke}, {Frail}, {Horesh}, {Corsi}, \&
			{Hotokezaka}}]{Mooley2018B}
		{Mooley} K.~P. {et~al.}, 2018{\natexlab{a}}, ArXiv e-prints
		
		\bibitem[{{Mooley} {et~al}\mbox{.}(2018{\natexlab{b}}){Mooley}, {Nakar},
			{Hotokezaka}, {Hallinan}, {Corsi}, {Frail}, {Horesh}, {Murphy}, {Lenc},
			{Kaplan}, {de}, {Dobie}, {Chandra}, {Deller}, {Gottlieb}, {Kasliwal},
			{Kulkarni}, {Myers}, {Nissanke}, {Piran}, {Lynch}, {Bhalerao}, {Bourke},
			{Bannister}, \& {Singer}}]{Mooley2018}
		{Mooley} K.~P. {et~al.}, 2018{\natexlab{b}}, \nat, 554, 207
		
		\bibitem[{{Nakar}(2007)}]{Nakar2007}
		{Nakar} E., 2007, \physrep, 442, 166
		
		\bibitem[{{Nakar} \& {Sari}(2012)}]{Nakar2012}
		{Nakar} E., {Sari} R., 2012, \apj, 747, 88
		
		\bibitem[{{Nava} {et~al}\mbox{.}(2012){Nava}, {Salvaterra}, {Ghirlanda},
			{Ghisellini}, {Campana}, {Covino}, {Cusumano}, {D'Avanzo}, {D'Elia},
			{Fugazza}, {Melandri}, {Sbarufatti}, {Vergani}, \& {Tagliaferri}}]{Nava2012}
		{Nava} L. {et~al.}, 2012, \mnras, 421, 1256
		
		\bibitem[{{Nava} {et~al}\mbox{.}(2014){Nava}, {Vianello}, {Omodei},
			{Ghisellini}, {Ghirlanda}, {Celotti}, {Longo}, {Desiante}, \& {Barniol
				Duran}}]{Nava2014}
		{Nava} L. {et~al.}, 2014, \mnras, 443, 3578
		
		\bibitem[{{Nysewander}, {Fruchter} \& {Pe'er}(2009){Nysewander}, {Fruchter}, \&
			{Pe'er}}]{Nysewander2009}
		{Nysewander} M., {Fruchter} A.~S., {Pe'er} A., 2009, \apj, 701, 824
		
		\bibitem[{{Oates} {et~al}\mbox{.}(2012){Oates}, {Page}, {De Pasquale},
			{Schady}, {Breeveld}, {Holland}, {Kuin}, \& {Marshall}}]{Oates2012}
		{Oates} S.~R., {Page} M.~J., {De Pasquale} M., {Schady} P., {Breeveld} A.~A.,
		{Holland} S.~T., {Kuin} N.~P.~M., {Marshall} F.~E., 2012, \mnras, 426, L86
		
		\bibitem[{{Panaitescu} \& {Kumar}(2001)}]{Panaitescu2001}
		{Panaitescu} A., {Kumar} P., 2001, \apjl, 560, L49
		
		\bibitem[{{Panaitescu} \& {M{\'e}sz{\'a}ros}(1999)}]{Panaitescu1999}
		{Panaitescu} A., {M{\'e}sz{\'a}ros} P., 1999, \apj, 526, 707
		
		\bibitem[{{Pescalli} {et~al}\mbox{.}(2015){Pescalli}, {Ghirlanda}, {Salafia},
			{Ghisellini}, {Nappo}, \& {Salvaterra}}]{Pescalli2015}
		{Pescalli} A., {Ghirlanda} G., {Salafia} O.~S., {Ghisellini} G., {Nappo} F.,
		{Salvaterra} R., 2015, \mnras, 447, 1911
		
		\bibitem[{{Racusin} {et~al}\mbox{.}(2016){Racusin}, {Oates}, {de Pasquale}, \&
			{Kocevski}}]{Racusin2016}
		{Racusin} J.~L., {Oates} S.~R., {de Pasquale} M., {Kocevski} D., 2016, \apj,
		826, 45
		
		\bibitem[{{Rhoads}(1999)}]{Rhoads1999}
		{Rhoads} J.~E., 1999, \apj, 525, 737
		
		\bibitem[{{Rossi}, {Lazzati} \& {Rees}(2002){Rossi}, {Lazzati}, \&
			{Rees}}]{Rossi2002}
		{Rossi} E., {Lazzati} D., {Rees} M.~J., 2002, \mnras, 332, 945
		
		\bibitem[{{Ruan} {et~al}\mbox{.}(2018){Ruan}, {Nynka}, {Haggard}, {Kalogera},
			\& {Evans}}]{Ruan2018}
		{Ruan} J.~J., {Nynka} M., {Haggard} D., {Kalogera} V., {Evans} P., 2018, \apjl,
		853, L4
		
		\bibitem[{{Ruderman}(1975)}]{Ruderman1975}
		{Ruderman} M., 1975, in Annals of the New York Academy of Sciences, Vol. 262,
		Seventh Texas Symposium on Relativistic Astrophysics, {Bergman} P.~G.,
		{Fenyves} E.~J., {Motz} L., eds., pp. 164--180
		
		\bibitem[{{Salafia} {et~al}\mbox{.}(2015){Salafia}, {Ghisellini}, {Pescalli},
			{Ghirlanda}, \& {Nappo}}]{Salafia2015}
		{Salafia} O.~S., {Ghisellini} G., {Pescalli} A., {Ghirlanda} G., {Nappo} F.,
		2015, \mnras, 450, 3549
		
		\bibitem[{{Salmonson}(2003)}]{Salmonson2003}
		{Salmonson} J.~D., 2003, \apj, 592, 1002
		
		\bibitem[{{Salvaterra} {et~al}\mbox{.}(2012){Salvaterra}, {Campana}, {Vergani},
			{Covino}, {D'Avanzo}, {Fugazza}, {Ghirlanda}, {Ghisellini}, {Melandri},
			{Nava}, {Sbarufatti}, {Flores}, {Piranomonte}, \&
			{Tagliaferri}}]{Salvaterra2012}
		{Salvaterra} R. {et~al.}, 2012, \apj, 749, 68
		
		\bibitem[{{Santana}, {Barniol Duran} \& {Kumar}(2014){Santana}, {Barniol
				Duran}, \& {Kumar}}]{Santana2014}
		{Santana} R., {Barniol Duran} R., {Kumar} P., 2014, \apj, 785, 29
		
		\bibitem[{{Sari}, {Piran} \& {Halpern}(1999){Sari}, {Piran}, \&
			{Halpern}}]{Sari1999}
		{Sari} R., {Piran} T., {Halpern} J.~P., 1999, \apjl, 519, L17
		
		\bibitem[{{Troja} {et~al}\mbox{.}(2017){Troja}, {Piro}, {van Eerten},
			{Wollaeger}, {Im}, {Fox}, {Butler}, {Cenko}, {Sakamoto}, {Fryer}, {Ricci},
			{Lien}, {Ryan}, {Korobkin}, {Lee}, {Burgess}, {Lee}, {Watson}, {Choi},
			{Covino}, {D'Avanzo}, {Fontes}, {Gonz{\'a}lez}, {Khandrika}, {Kim}, {Kim},
			{Lee}, {Lee}, {Kutyrev}, {Lim}, {S{\'a}nchez-Ram{\'{\i}}rez}, {Veilleux},
			{Wieringa}, \& {Yoon}}]{Troja2017}
		{Troja} E. {et~al.}, 2017, \nat, 551, 71
		
		\bibitem[{{van der Horst} {et~al}\mbox{.}(2014){van der Horst}, {Paragi}, {de
				Bruyn}, {Granot}, {Kouveliotou}, {Wiersema}, {Starling}, {Curran}, {Wijers},
			{Rowlinson}, {Anderson}, {Fender}, {Yang}, \& {Strom}}]{VanderHorst2014}
		{van der Horst} A.~J. {et~al.}, 2014, \mnras, 444, 3151
		
		\bibitem[{{van Eerten} \& {MacFadyen}(2012)}]{VanEerten2012}
		{van Eerten} H.~J., {MacFadyen} A.~I., 2012, \apj, 751, 155
		
		\bibitem[{{Wanderman} \& {Piran}(2010)}]{Wanderman2010}
		{Wanderman} D., {Piran} T., 2010, \mnras, 406, 1944
		
		\bibitem[{{Wygoda} {et~al}\mbox{.}(2016){Wygoda}, {Guetta}, {Mandich}, \&
			{Waxman}}]{Wygoda2016}
		{Wygoda} N., {Guetta} D., {Mandich} M.~A., {Waxman} E., 2016, \apj, 824, 127
		
		\bibitem[{{Zhang} \& {M{\'e}sz{\'a}ros}(2002)}]{Zhang2002}
		{Zhang} B., {M{\'e}sz{\'a}ros} P., 2002, \apj, 571, 876
		
		\bibitem[{{Zhang} {et~al}\mbox{.}(2015){Zhang}, {van Eerten}, {Burrows},
			{Ryan}, {Evans}, {Racusin}, {Troja}, \& {MacFadyen}}]{Zhang2015}
		{Zhang} B.-B., {van Eerten} H., {Burrows} D.~N., {Ryan} G.~S., {Evans} P.~A.,
		{Racusin} J.~L., {Troja} E., {MacFadyen} A., 2015, \apj, 806, 15
		
	\end{thebibliography}
\end{document}